%% file: ms.tex
\documentclass{ieeeaccess}
\usepackage{cite}
\usepackage{amsmath,amssymb,amsfonts}
\usepackage{algorithmic}
\usepackage{graphicx}
\usepackage{textcomp}
\def\BibTeX{{\rm B\kern-.05em{\sc i\kern-.025em b}\kern-.08em
    T\kern-.1667em\lower.7ex\hbox{E}\kern-.125emX}}

\ifCLASSOPTIONcompsoc
\usepackage[caption=false, font=normalsize, labelfont=sf, textfont=sf]{subfig}
\else
\usepackage[caption=false, font=footnotesize]{subfig}

\usepackage{xspace}
\usepackage{threeparttable}
\usepackage{cleveref}
\usepackage{booktabs} % For formal tables
\usepackage{multirow}

\usepackage{amsmath}

\usepackage{url}

\usepackage{color,soul} % For commenting & highlighting

\usepackage{comment}

\usepackage[shortlabels]{enumitem}

\newcommand{\hltext}{0} % For selecting highlighted text

\newcommand\blfootnote[1]{%
	\begingroup
	\renewcommand\thefootnote{}\footnote{#1}%
	\addtocounter{footnote}{-1}%
	\endgroup
}

\begin{document}
\history{Date of publication xxxx 00, 0000, date of current version xxxx 00, 0000.}
\doi{10.1109/ACCESS.2022.3181630}

\title{Improving the Security of the IEEE 802.15.6 Standard for Medical BANs}
\author{\uppercase{Muhammad Ali Siddiqi}\authorrefmark{1,2}, \uppercase{Georg Hahn}\authorrefmark{3}, \uppercase{Said Hamdioui}\authorrefmark{1}, \IEEEmembership{Senior Member, IEEE}, \uppercase{Wouter A. Serdijn}\authorrefmark{3},
	\IEEEmembership{Fellow, IEEE}, \uppercase{and Christos Strydis}\authorrefmark{1,2}, \IEEEmembership{Senior Member, IEEE}}
\address[1]{Quantum \& Computer Engineering department, Delft University of Technology, The Netherlands}
\address[2]{Department of Neuroscience, Erasmus Medical Center, Rotterdam, The Netherlands}
\address[3]{Section Bioelectronics, Delft University of Technology, The Netherlands}

\tfootnote{This work was supported by the EU-Funded Projects SDK4ED (Gr. Agr. No. 780572) and EuroEXA (Gr. Agr. No. 754337).}

\markboth
{Siddiqi \headeretal: Improving the Security of the IEEE 802.15.6 Standard for Medical BANs}
{Siddiqi \headeretal: Improving the Security of the IEEE 802.15.6 Standard for Medical BANs}

\corresp{Corresponding authors: Muhammad Ali Siddiqi (e-mail: m.a.siddiqi@tudelft.nl); Christos Strydis (e-mail: c.strydis@erasmusmc.nl).}

\begin{abstract}
A Medical Body Area Network (MBAN) is an ensemble of collaborating, potentially heterogeneous, medical devices located inside, on the surface of or around the human body with the objective of tackling one or multiple medical conditions of the MBAN host. These devices -- which are a special category of Wireless Body Area Networks (WBANs) -- collect, process and transfer medical data outside of the network, while in some cases they also administer medical treatment autonomously. Since communication is so pivotal to their operation, the newfangled IEEE 802.15.6 standard is aimed at the communication aspects of WBANs. It places a set of physical and communication constraints while it also includes association/disassociation protocols and security services that WBAN applications need to comply with. However, the security specifications put forward by the standard can be easily shown to be insufficient when considering realistic MBAN use cases and need further enhancements. The present work addresses these shortcomings by, first, providing a structured analysis of the IEEE 802.15.6 security features and, afterwards, proposing comprehensive and tangible recommendations on improving the standard's security.
\end{abstract}

\begin{keywords}
Wireless body area network, WBAN, MBAN, IEEE 802.15.6, Implantable medical device, IMD, Security 
\end{keywords}

\titlepgskip=-15pt

\maketitle

\blfootnote{This article has been accepted for publication in a future issue of this journal, but has not been fully edited. Content may change prior to final publication. Citation information: DOI 10.1109/ACCESS.2022.3181630, IEEE Access. This work is licensed under a Creative Commons Attribution 4.0 License. For more information, see https://creativecommons.org/licenses/by/4.0/.}

\input{paperbody}

%\appendices
%Appendixes, if needed, appear before the acknowledgment.

\bibliographystyle{IEEEtran}
\bibliography{101_References}

\begin{IEEEbiography}[{\includegraphics[width=1in,height=1.25in,clip,keepaspectratio]{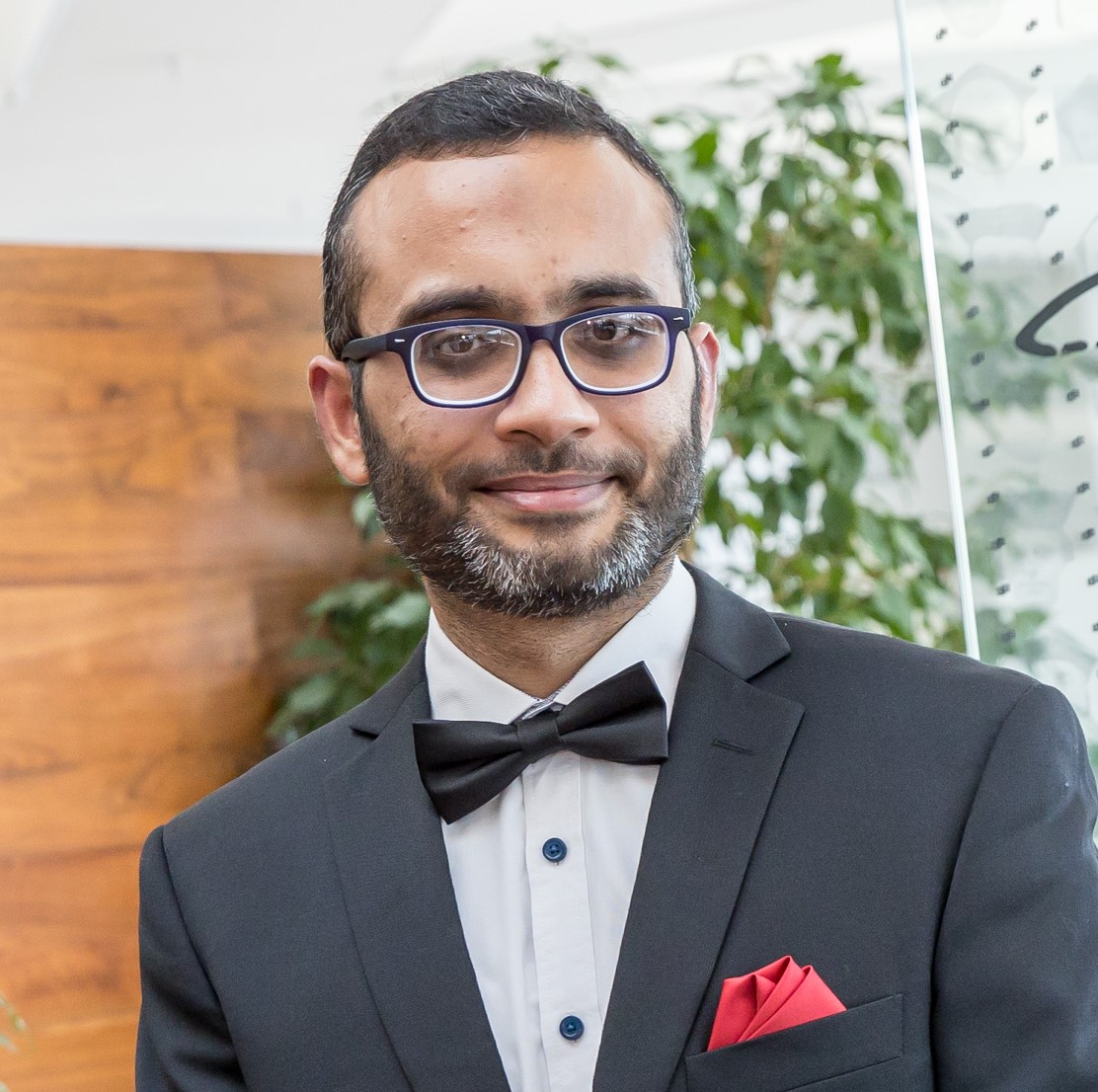}}]{Muhammad Ali Siddiqi} received the B.E. degree in electrical engineering from National University of Sciences and Technology, Islamabad, Pakistan, in 2009 and the (joint) M.Sc. degree in embedded computing systems from Norwegian University of Science and Technology, Trondheim, Norway, and University of Southampton, UK, in 2012. In 2021, he received the Ph.D. degree from the Erasmus University Rotterdam, the Netherlands, after the culmination of his research on the security and privacy aspects of IMDs. Currently, he is pursuing postdoctoral research at the Computer Engineering Laboratory of the Delft University of Technology, the Netherlands. From 2012 to 2017, he worked as a Design Engineer at Silicon Labs Norway on the ultra-low-power MCU design. His research interest includes the development of security protocols and computing architectures for heavily resource-constrained embedded systems, such as IMDs.
\end{IEEEbiography}

\begin{IEEEbiography}[{\includegraphics[width=1in,height=1.25in,clip,keepaspectratio]{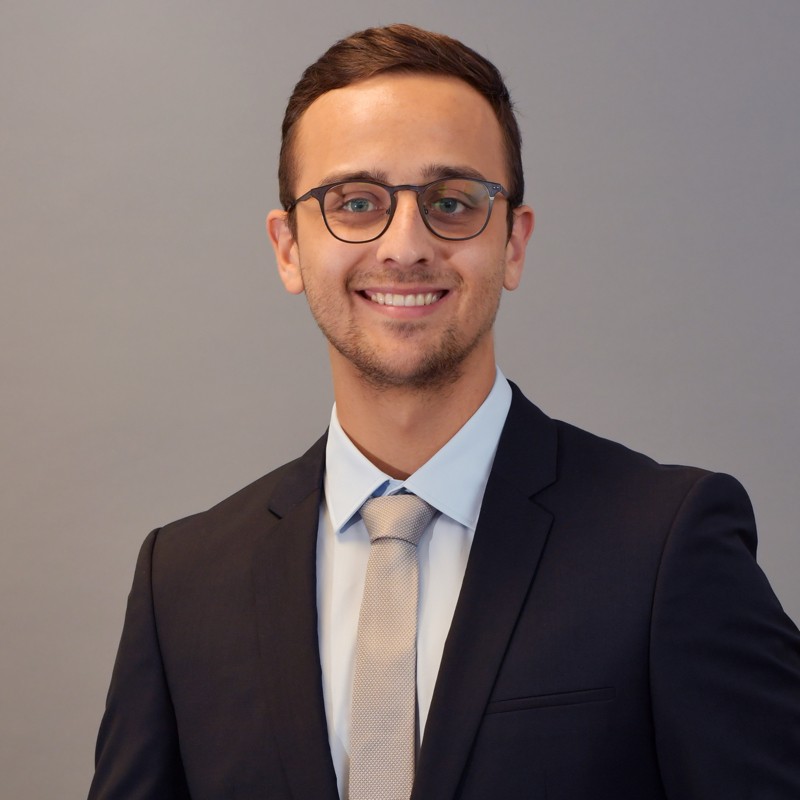}}]{Georg Hahn} received the B.Sc. degree in electrical engineering and information technology from the University of Technology Vienna, Austria, in 2019 and the M.Sc. degree in biomedical engineering from the Delft University of Technology, the Netherlands, in 2021. He pursued his master’s thesis with the Neuroscience Department of the Erasmus Medical Center in the Netherlands. His research interests include cybersecurity of medical devices and medical networks, as well as microelectronic systems.
Currently, he is working as a technology consultant for a global consulting firm based in Vienna, Austria.
\end{IEEEbiography}

\begin{IEEEbiography}[{\includegraphics[width=1in,height=1.25in,clip,keepaspectratio]{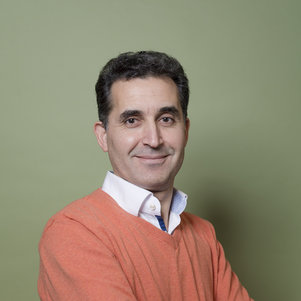}}]{Said Hamdioui} (Senior Member, IEEE) is currently Chair Professor on Dependable and Emerging Computer Technologies, Head of the Quantum and Computer Engineering department, and also serving as Head of the Computer Engineering Laboratory (CE-Lab) of the Delft University of Technology, the Netherlands. He received the MSEE and PhD degrees (both with honors) from TU Delft. Prior to joining TU Delft as a professor, Hamdioui spent over seven years within industry including Microprocessor Products Group at Intel Corporation (California, USA), IP and Yield Group at Philips Semiconductors R\&D (Crolles, France) and DSP design group at Philips/NXP Semiconductors (Nijmegen, The Netherlands). His research focuses on two domains: emerging technologies and computing paradigms, and hardware dependability.	
\end{IEEEbiography}

\begin{IEEEbiography}[{\includegraphics[width=1in,height=1.25in,clip,keepaspectratio]{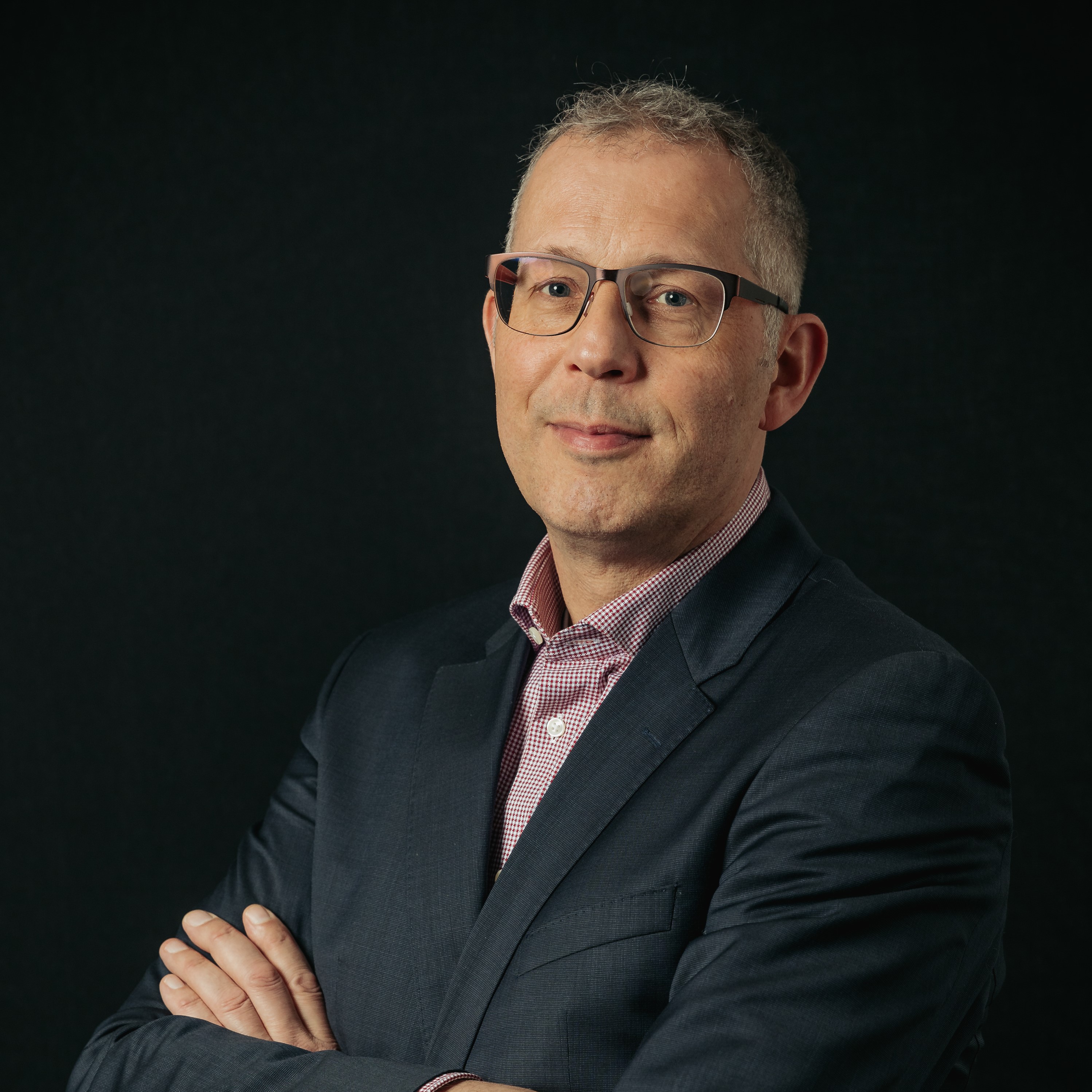}}]{Wouter A. Serdijn} (Fellow, IEEE) was born in Zoetermeer ('Sweet Lake City'), the Netherlands, in 1966. He received the M.Sc. (cum laude) and Ph.D. degrees from Delft University of Technology, Delft, The Netherlands, in 1989 and 1994, respectively. Currently, he is a full professor in bioelectronics at Delft University of Technology, where he heads the Section Bioelectronics, and a visiting honorary professor at University College London, in the Analog and Biomedical Electronics group. His research interests include integrated biomedical circuits and systems for biosignal conditioning and detection, neuroprosthetics, transcutaneous wireless communication, power management and energy harvesting as applied in, e.g., cardiac pacemakers, cochlear implants, neurostimulators, bioelectronic medicine and electroceuticals.	
\end{IEEEbiography}

\begin{IEEEbiography}[{\includegraphics[width=1in,height=1.25in,clip,keepaspectratio]{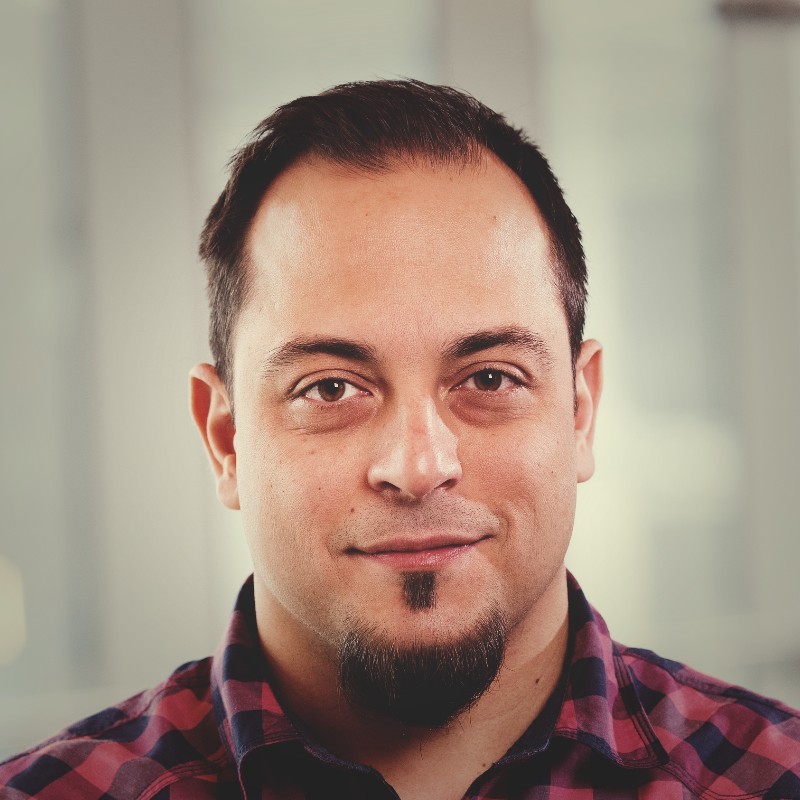}}]{Christos Strydis} (Senior Member, IEEE) received his M.Sc. \textit{(magna cum laude)} and Ph.D. degrees in computer engineering from the Delft University of Technology. He holds a dual Associate-Professor position with the Neuroscience department of the Erasmus Medical Center and with the Quantum \& Computer Engineering department of the Delft University of Technology. He is also Head of the Neurocomputing Laboratory at the Erasmus Medical Center. He has published work in well-known international conferences and journals. He has delivered invited talks in various venues. His current research interests include brain simulations, high-performance computing, low-power embedded (implantable) systems, and functional ultrasound imaging.
\end{IEEEbiography}

\EOD

\end{document}

%% file: paperbody.tex
\input{01_Introduction}

\input{02_WBAN_background}

\input{03_MBAN_security_concerns}

\input{04_The_IEEE_802.15.6_standard}

\input{05_Exploring_future_MBAN_application_scenarios}

\input{08_Conclusion}

%% file: 01_Introduction.tex
\section{Introduction}
\label{sec:introduction}

The recent shift from stationary, offline devices to interconnected, smart electronics and sensors catalyzed a wide range of novel technologies and applications. The so-called \textit{Internet of Things} (IoT) has become one of the most disruptive technologies of our decade, enabling the creation of device networks, which try to create value in nearly every industry. Especially the healthcare sector is undergoing a large technological transformation. \textit{Medical Body Area Networks} (MBANs) are revolutionizing the way in which healthcare data is gathered and processed by creating a network of interconnected nodes inside, on the surface or in the vicinity of the human body. Nodes can comprise a variety of devices, e.g., medical implants, sensors and wearables that are used to measure or regulate diverse biomedical signals, e.g., respiratory rate, mechanical motion, heart rate and so on. This facilitates multiple interesting applications in areas such as real-time health monitoring, ambient-assisted living (AAL) and pathology treatment.

Although MBANs show a very high potential for future applications, increased functionality is always accompanied by a proportional increase in potential risks. Given the sensitivity of the data processed by the nodes and the critical functionality of the network's actuators, a security-conscious implementation is crucial. Cybersecurity attacks, such as \textit{Denial of Service} (DoS), can have life-threatening consequences for the patient and must be prevented. However, not only threats to the patient's life are an issue but the patient's privacy is also at stake. MBANs are typically handling highly sensitive \textit{Personal Health Information} (PHI), for which confidentiality, integrity and availability need to be ensured. Due to the limited resources (e.g., memory, battery life etc.) offered by implanted or wearable nodes, this can often become a challenging task. Although some modern nodes manage to use state-of-the-art security implementations and protocols, there are still a number of attack vectors enabled through wireless connections and vulnerabilities.

\begin{figure*}[!t]
\centering
  \includegraphics[trim={0cm 1.2cm 0cm 1.2cm},clip,scale=0.48]{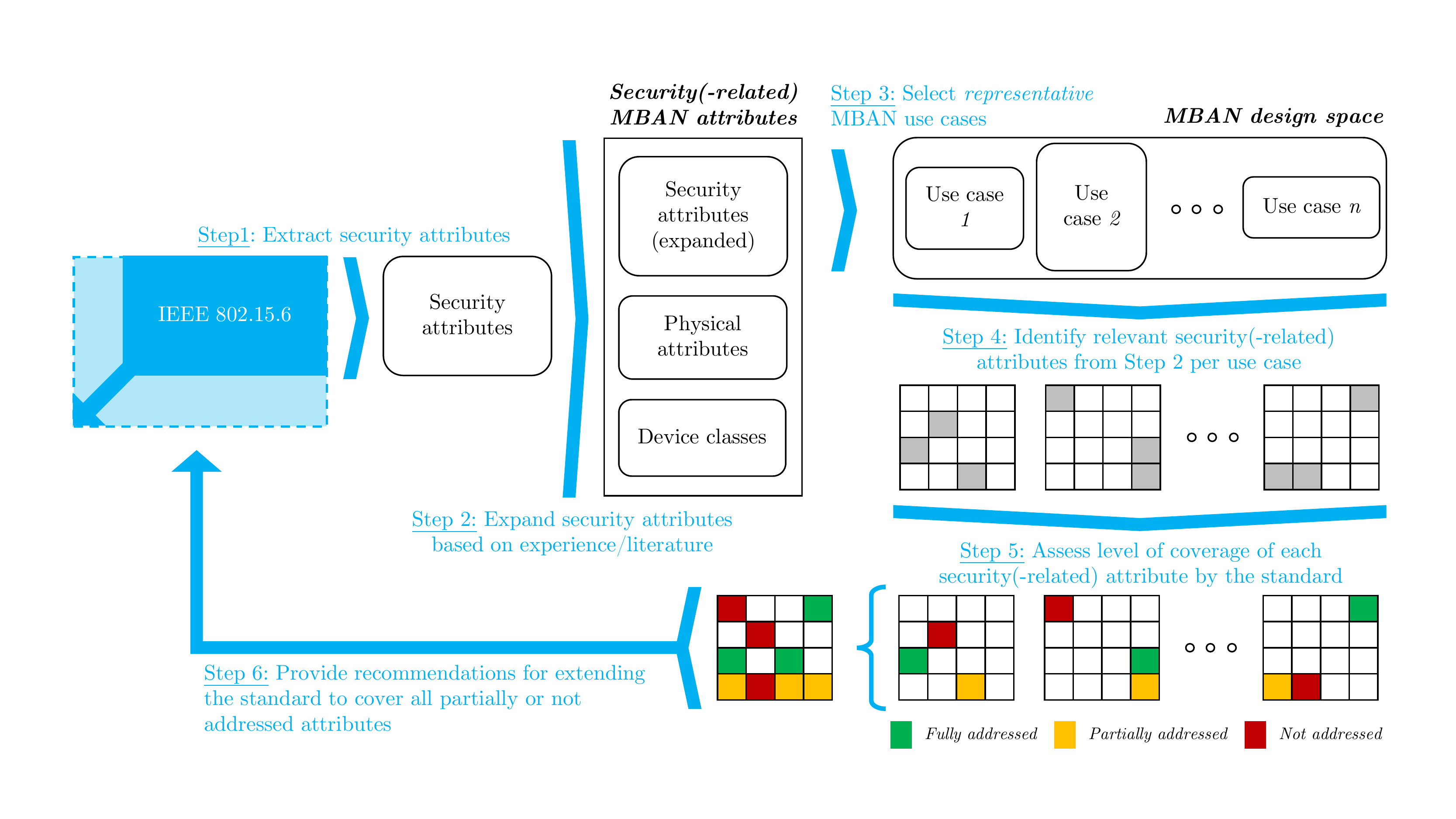}
  \caption{Overview of the methodology employed in this work to improve the security of IEEE 802.15.6.
    \ifnum\hltext>0
    \hl{The choice of the specific locations of the security(-related) attributes in steps 4 and 5 is just as an example.}
    \else
    The choice of the specific locations of the security(-related) attributes in steps 4 and 5 is just as an example.
    \fi
  }
  \label{fig:mban-method}
\end{figure*}

In 2007, a new task group (TG6) was created within the IEEE 802.15 (IEEE 802 WG15) family to satisfy the requirements of such devices.  
In 2010, the IEEE 802.15 TG6, or IEEE 802.15.6, published the first draft of a standard that was optimised for low-power nodes for medical and non-medical applications, which was then approved and ratified in 2012~\cite{ieee_standard}.
IEEE 802.15.6 aims to govern WBAN communication at the PHY and MAC layer level, and to provide several security measures and protocols.
\ifnum\hltext>0
\hl{The standard provides three security paradigms for message protection ranging from unsecured communication for non-critical applications to encrypted communication with authentication for critical applications.}
\else
The standard provides three security paradigms for message protection ranging from unsecured communication for non-critical applications to encrypted communication with authentication for critical applications.
\fi
However, as with any novel standard or technology, security issues have been found~\cite{toorani2016security}.
This is compounded by the fact that much has changed in the WBAN landscape (in literature) since 2012, and many new medical applications that fall under the category of WBANs (such as Neural Dust) do not exactly fit within the framework set by 802.15.6.
This work\footnote{This work is derived from the master's thesis of Georg Hahn~\cite{hahn2021assessing} that was developed under the supervision of Dr. Muhammad Ali Siddiqi, Prof. Wouter A. Serdijn and Dr. Christos Strydis.}, thus, takes a different approach to tackle the security problem of this standard. Instead of finding vulnerabilities in the employed protocols and primitives, we take a step back and analyse the standard through the lens of \textit{realistic} use cases that were unfortunately not considered at the time the standard was drafted.
The aim of this work is, thus, \textit{to improve the security posture of the IEEE 802.15.6 standard using representative MBAN use cases}. 

A schematic overview of the methodology we follow in this work is provided in Figure~\ref{fig:mban-method}. First, the security attributes included in the standard are extracted. It should be noted that this extraction is based on the authors' own estimations, since these attributes are not explicitly identified in the standard. These security attributes are, then, extended initially with additional security attributes available in cybersecurity literature, and augmented further with miscellaneous attributes to include MBAN-device classes and physical attributes. These latter ones are not security attributes, strictly speaking, but indirectly affect the former ones and are thus cumulatively termed security-related attributes in this manuscript.

Although it is imperative to cover all possible aspects of MBAN security, alas, the combinatorics of the extended set of all these security and security-related attributes results in a vast search space of potential MBAN system configurations. Attempting to extend the IEEE standard so as to address all possible MBAN configurations in the projected solution space is intractable. This work, thus, relies on the insight that not all possible MBAN configurations yield meaningful designs. Instead, the careful selection of an arbitrarily small but \textit{representative} subset of MBAN use cases can \textit{sufficiently} cover the design (sub)space of realistic solutions by making sure that the combined specifications of all selected use cases address all security(-related) attributes identified in the previous step (shown as grayed table elements in Figure~\ref{fig:mban-method}). The standard can, then, be validated against these select use cases: if a shortcoming is found, it can be formulated as a recommendation for improving the standard. The working details of this methodology will be further clarified through the rest of the paper.  Suffice to say at this point that our methodology assumes an iterative, refining process with the inclusion of more use cases or security vulnerabilities in the future, which is reflected in the closed loop depicted in Figure~\ref{fig:mban-method}.

Overall, the contributions of this work are as follows:

\begin{itemize}
    \item An overview of the IEEE 802.15.6 standard with a focus on the security features and vulnerabilities is given.
    \item A structured security-assessment methodology for exhaustively analyzing the standard is put forward. It entails defining realistic MBAN use cases to deduce specific security and physical attributes.
    \item Using said methodology, the standard's security posture is assessed and problems are identified.
    \item Based on the assessment findings, specific recommendations are provided for improving IEEE 802.15.6.
\end{itemize}

The rest of the paper is organized as follows: In Section~\ref{sec:WBAN_background}, we describe the MBAN concept in detail. In Section~\ref{chap:mban_security_requirements_and_threat_landscape}, we discuss their security attributes and the threat landscape. Section~\ref{sec:IEEE_802.15.6_security} introduces the IEEE 802.15.6 standard from the perspective of security. In Section~\ref{chap:assesing_and_analysing_the_standard}, a structured process to assess the security posture of this standard is introduced and used. Based on this assessment, we proceed to also offer recommendations for improving the standard in terms of security. We draw overall conclusions in Section~\ref{chap:conclusion}.

%% file: 02_WBAN_background.tex
\section{MBAN background}
\label{sec:WBAN_background}

MBANs, sometimes also referred to as the Internet of Medical Things (IoMT), are a special case of Wireless Body Area Networks (WBANs), which generally are a subgroup of Wireless Sensor Networks (WSNs). By comparison to the other categories, WSNs comprise a large set of sensors with high computational resources dispersed in the environment. These sensors often collect relevant data of their environmental conditions (e.g., traffic levels in cities, manufacturing flow in factories, air toxicity in chemical plants, etc.). While WBANs also consist of numerous wirelessly connected sensors and actuators, they are specifically placed in, on or around the human body. Sensors and actuators in such networks very often do not have full computational capabilities, especially when implanted. The aim of these sensors is to gather biological data, which is then transmitted to a server, where more resource-intensive computations (e.g., data analysis, machine learning algorithms, predictive analytics, etc.) are carried out. If needed, the server can also decide to initiate an action and send commands to the actuator nodes via a central coordinator. If the collected data is used for medical purposes, the network is called an MBAN.
The exact differences between WSNs and WBANs are listed in Table~\ref{tab:wsn_vs_wban}.

Relying on our expertise in the medical domain, we focus specifically on MBANs in this paper without loss of generality.
However, our methodology is applicable to the broader WBAN category as well. 
In the ongoing amendment to the IEEE 802.15.6 standard (i.e., IEEE 802.15.6a)~\cite{IEEE:2022}, the scope of a WBAN is not constrained to humans anymore as it is categorized into Human BAN (HBAN) and Vehicle BAN (VBAN).
And so, this work is primarily aligned with the HBAN category.

\begin{table}[!t]
    \centering
    \caption{Detailed comparison between WSNs and WBANs~\cite{Qu:2019}}
    \begin{tabular}{lll}
        \hline
        \textbf{Metric}  & \textbf{WSN}            & \textbf{WBAN}          \\ \hline
        Range             & m to km                 & cm to m               \\
        No. of nodes    & 100s                & <10                   \\
        Node size         & No special requirements & Very small            \\
        Node task         & Single or scheduled     & Multiple              \\
        Network topology  & Fixed                   & Variable              \\
        Data loss         & Tolerable               & Intolerable           \\
        Node placement    & Relatively easy         & Difficult             \\
        Bio-compatibility & -                       & Critical              \\
        Node life         & Months/years            & The longer the better \\
        Safety            & Relatively less critical  & Critical              \\
        Security          & Relatively less critical          & Critical         \\
        Standard          & IEEE 802.11.4           & IEEE 802.15.6         \\ \hline
    \end{tabular}
    \label{tab:wsn_vs_wban}
\end{table}

\subsection{Node types}
\label{sec:node_types}

The wide variety of use cases for MBANs inherits a need for a multitude of different nodes with different requirements and challenges. Nodes mostly act as \textit{autonomous} devices and they need to be fully equipped with a communication system to relay data to other nodes inside the network or to the outside world~\cite{Pramanik:2019}. They can be classified on the basis of their (a) functionality, (b) type of implementation, and (c) specific role in the network, as summarised in Table~\ref{tab:node_classification}.

\begin{table}[!t]
    \centering
    \caption{Classification of MBAN nodes according to functionality, implementation style or special role}
    \label{tab:node_classification}
    \begin{tabular}{lll}
        \hline
        Functionality                    & Implementation & Role             \\ \hline
        Sensor                           & Invasive       & End node         \\
        Actuator                         & Semi-invasive  & Relay node       \\
        Hybrid                           & Wearable       & Coordinator node \\
        Central Control Unit (CCU)       & Ambient        &                  \\
        \hline
    \end{tabular}
\end{table}

\subsubsection{Functionality}

Depending on their functionality, the different node types are described as follows:

The main task of a \textbf{sensor} node is to gather relevant data and transmit it to another node or a coordinator. Due to their very limited memory capacity, they need to transmit data in a specified time interval in order to mitigate the risk of memory overflow, thereby losing data. In addition, their energy-storage capacities are also scarce, making it crucially important to design effective protocols and processes.

An \textbf{actuator} node's main task is to perform an intervention on the human body based on the information it receives from other nodes. This action can entail releasing some drug in the body or even stimulating certain areas of the brain. As it is sometimes challenging to recharge actuator nodes, the action needs to be carried out in an energy-efficient manner.

A \textbf{hybrid} node has both sensing and actuating capabilities. The most prominent representatives of this category are implantable medical devices (IMDs). For example, an implantable cardiac defibrillator (ICD) can both sense cardiac arrhythmia and deliver shocks to treat the condition.

The \textbf{central-control unit (CCU)}, sometimes also called the central coordinator, personal control unit or base station, is the main point of coordination. It collects the data sent by sensor nodes, transmits signals to actuator nodes and is responsible for relaying the gathered information to beyond-MBAN communication channels. Given its energy-consuming and CPU-intensive tasks, the CCU usually possesses considerably more battery capacity, memory size and processing power than other nodes.

\subsubsection{Implementation}
\label{sec:implementation}

An MBAN node can belong to one of the following categories based on the relative proximity to the human body:

\begin{itemize}
    \item \textbf{Invasive:} Nodes in this category are implanted under the skin or within the human body.

    \item \textbf{Semi-invasive:} Here, a part of the node is implanted, and the rest is outside the human body.

    \item \textbf{Wearable:} These nodes are located on the human body and most often need to have direct contact with the skin in order to function properly.

    \item \textbf{Ambient:} Nodes that are in close proximity, surrounding the body, are part of this category.

\end{itemize}

\subsubsection{Role}

Based on the function or role in the network, nodes can be classified into an \textbf{end node} (which is at the end of a communication line), a \textbf{relay node} (which receives data and relays it to another node) or a \textbf{coordinator node} (which orchestrates the routing procedures). Relay nodes help in increasing the communication distance while simultaneously reducing the transmission-energy costs.

Some example nodes classified by the introduced classifiers can be seen in Table~\ref{tab:node_classification_examples}.

\begin{table*}[!t]
    \centering
    \caption{Classification of several example nodes found in an MBAN}
    \label{tab:node_classification_examples}
    \footnotesize
    \begin{tabular}{cccc}
        \hline
        \multicolumn{1}{c}{Nodes} & \multicolumn{1}{c}{Functionality} & \multicolumn{1}{c}{Implementation} & \multicolumn{1}{c}{Role} \\ \hline
        Heart-rate sensor       & Sensor   & Wearable            & End node / Relay node \\
        Insulin pump            & Actuator & Semi-invasive       & End node              \\
        Cochlear Implant        & Hybrid   & Semi-invasive       & End node              \\
        Mobile phone            & CCU      & Ambient             & Coordinator node      \\
        ECG / EMG / EEG         & Sensor   & Wearable / Invasive & End node / Relay node \\
        Endoscope capsule       & Sensor   & Invasive            & End node              \\
        Gyroscope               & Sensor   & Wearable            & End node / Relay node \\
        Pacemaker               & Hybrid   & Invasive            & End node / Relay node \\
        Smart watch             & CCU      & Wearable            & Coordinator node      \\
        Stimulation electrode   & Actuator & (Semi-)invasive     & End node              \\
        Vagus-nerve stimulation & Hybrid   & Invasive            & End node / Relay node \\
        \hline
    \end{tabular}
\end{table*}

\subsection{MBAN ecosystem}
\label{sec:WBAN_architecture}

The MBAN ecosystem consists of different communication layers that can be split up into three tiers~\cite{Roy:2018}:
\begin{itemize}
    \item Tier 1: Intra-MBAN communication
    \item Tier 2: Inter-MBAN communication
    \item Tier 3: Beyond-MBAN communication
\end{itemize}

\begin{figure*}[!t]
  \centering
  \includegraphics[scale=0.8]{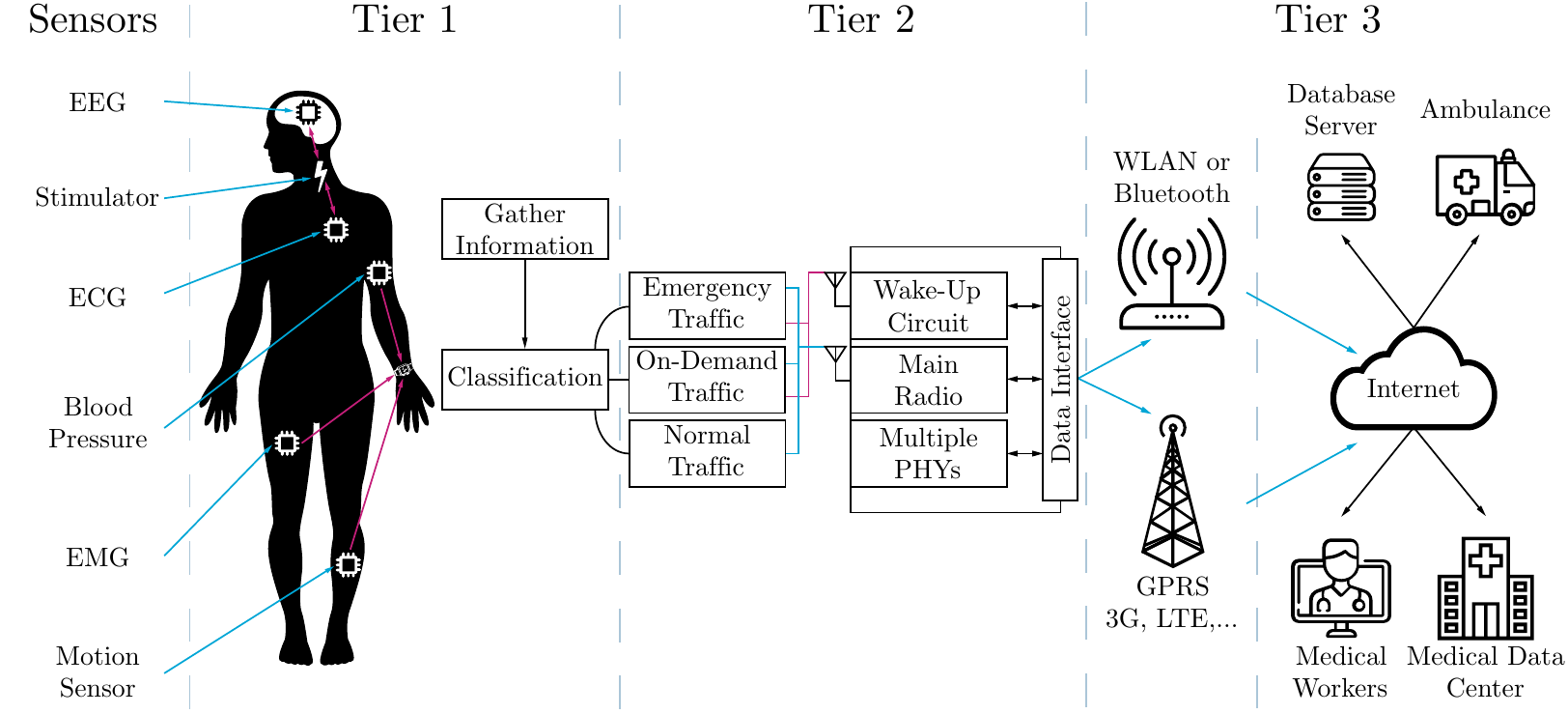}
  \caption{A detailed overview of a WBAN ecosystem for medical and non-medical applications, based on~\cite{Ullah:2012}}
  \label{fig:detailed_architecture_of_WBAN}
\end{figure*}

As shown in Figure~\ref{fig:detailed_architecture_of_WBAN}, in Tier 1 the different sensors and actuators gather biological data, e.g., blood pressure, ECG, EEG etc., and transmit it -- depending on the network topology -- to a collector node where the data is classified. In Tier 2, the collected data gets transmitted to a coordinator acting as a sink. This coordinator can be a smart phone, computer or some other personal-communication device that classifies the data and subsequently transfers it via WLAN, GPRS etc. to remote servers in Tier 3 (e.g., Cloud infrastructure).

These servers collect and analyze the data and provide it to a medical professional who, in turn, performs medical diagnostics, or the server directly calculates the next action using technologies such as machine learning. In the case of receiving abnormal data, Tier-3 endpoints will carry out emergency responses to speed up the process~\cite{Qu:2019}. Additionally, if the employed nodes have enough processing capabilities, data could be processed in a closed loop in Tiers 1 and 2. This could be useful for highly time-critical calculations or when the network is not connected to a Tier-3 Cloud server.

\subsection{Topologies}
\label{sec:topologies}

Depending on the application, the sensor and actuator nodes in Tier 1 can be connected in different topologies: \textit{star, tree} or \textit{peer to peer}.

Amongst these, the most popular is the star, as this arrangement does not need a routing protocol, making the delay between data packets minimal. Here, a data collector acts as base station to which all sensor and actuator nodes are connected. This central coordinator often has superior energy and computational capacities than other nodes since managing the entire network is a resource-intensive task. While the simplicity of this topology yields great advantages, the central coordinator represents a \textit{single point of failure} (SPOF), which
\ifnum\hltext>0
\hl{should be avoided}
\else
should be avoided
\fi
for high availability systems like MBANs. A use case of this approach is given in~\cite{Pan:2014}, where a wireless, MBAN-based 3-lead ECG is realised.

The tree topology, similar to the star, is governed by a single root node at the top of the structure. It can also be seen as a multi-hop star topology, in which the root node acts as the coordinator and the branching sensor and actuator nodes collect the data. Although this topology offers great flexibility and, due to the subordinate relay nodes, higher reliability, the root node also is an SPOF. Kim et al.~\cite{Kim:2016} proposed such a WBAN configuration using a multi-hop tree topology.

In a peer-to-peer arrangement, sensor nodes in a radio range are either directly communicating to each other or communicating through multiple hops. This type of topology does not rely on an upstream data-transfer, but instead transfers collected data from node to node without paying attention to the network's hierarchical structure. This topology trades energy efficiency and battery lifetime for increased reliability, since in case of a node failure the communication can be re-routed. However, routing protocols become increasingly more complex and are often difficult to realise.
Given the inherent complexity, this topology is not as commonly used as previous configurations.

\subsection{MBAN applications}
\label{sec:MBAN_applications}

Having discussed all the necessary details to understand the concept of MBANs, some potential applications\footnote{
\ifnum\hltext>0
\hl{It should be noted that despite some similarities between the two, \textit{implementation} (from Section~{\ref{sec:implementation}}) refers to any MBAN \textit{node}, whereas \textit{applications} in this section refer to the potential medical use cases.}
\else
It should be noted that despite some similarities between the two, \textit{implementation} (from Section~\ref{sec:implementation}) refers to any MBAN \textit{node}, whereas \textit{applications} in this section refer to the potential medical use cases.
\fi
} will be presented next, which can be further divided into four categories: implantable, semi-invasive, wearable and remote control of medical devices.

\subsubsection{Implantable}

In this category, sensor and actuator nodes are usually implanted inside the body, beneath the skin or in the blood stream. Here are some example applications:

\textbf{Cardiac diseases:} A common procedure to prevent arrhythmia in high-risk patients, is to implant a cardiac pacemaker that autonomously delivers shocks whenever needed. However, traditional pacemakers are rather chunky and invasive, and implanting the leads can be challenging. Therefore, next-generation leadless cardiac pacemakers (LCPs) use MBAN technology to miniaturise the form factor (up to 80\% smaller) and reduce invasiveness of the application. The self-contained electrode system of the LCP gets implanted directly into the right ventricle, eliminating several complications associated with traditional pacemakers, such as lead fracture and pocket infections~\cite{Groner:2019}. With the use of MBANs, biological signals can also be gathered and analysed to prevent myocardial infarction. Thereby multiple sensor nodes monitor the patients' vital signs, which get sent to a remote monitoring centre, where it is decided if therapy needs to be delivered~\cite{Hadjem:2013, Hadjem:2014}.

\textbf{Cancer detection:} A network of sensor nodes can be implemented to detect cancer, for instance by measuring the amount of nitric oxide included in cancerous cells. By monitoring related data, doctors can potentially diagnose tumours without the need of biopsy, providing more rapid analysis and treatment~\cite{Movassaghi:2014}.

\subsubsection{Semi-invasive}

Applications in this category can have a mixture of implanted and wearable sensor and actuator nodes. A possible application in this domain is \textbf{Diabetes control}. The golden standard for glucose monitoring is to self-monitor blood glucose levels after taking a small blood sample by pricking the finger. This method is rather invasive and inconvenient for the patient~\cite{Huzooree:2019}. Therefore, \textit{continuous glucose-monitoring systems} (CGMS) have been realised. Here, a glucose-sensor node is implanted into the body. The monitored blood-glucose levels are then sent to an insulin pump, which decides if actions need to be taken. There are already a variety of commercial solutions available, such as the DexCom G5 and the Medtronic MiniMed 670G. To the best of our knowledge, such insulin pumps, which comprise at most 3 nodes, are the only MBANs deployed commercially at the moment.

\subsubsection{Wearable}

Here, sensor nodes are usually attached to the skin using straps or worn by the patient in the form of a fabric, wristband, headgear etc. Some applications of this approach are:

\textbf{Epileptic-seizure detection:} Traditional, wired methods of epilepsy detection are not adequate for long-time monitoring without restricting patient mobility. Using the MBAN technology, a real-time monitoring system to detect and possibly predict incoming seizures can be realised. Escobar Cruz et al.~\cite{Escobar:2018} proposed a system to detect tonic-clonic seizures. A wearable glove with sensor nodes collects ECG signals and sends them to the patient's phone, which in turn communicates with a cloud computing server. With the help of a support vector machine, which is a supervised learning model, abnormal signals are recognised and detected. If a seizure is imminent, an automatic SMS message gets sent to a medical professional or a relative.

\textbf{Mental-status monitoring:} The wide variety of physiological signals collected by modern wearable sensors enable a multitude of different diagnostic possibilities. One of them is to use those signals to detect the mental health status of patients, of which stress monitoring is the most common~\cite{Wu:2019}. Audio and heart rate signals fed into machine learning algorithms can potentially detect mental stress levels in children, creating the possibility to remotely monitor child safety~\cite{Choi:2017}. Other ways to use the collected data include suicide risk monitoring~\cite{Alam:2014} and monitoring the state of mental health in chronically depressed patients~\cite{Saha:2016}.

\textbf{Sleep analysis:} Healthy sleep is essential in order to maintain both physical and mental health. Conditions like sleep apnea and insomnia can cause severe damage to an individual's well-being, if left untreated.
To monitor the sleep cycle a technique called \textit{polysomnography} (PSG) is used. Here, data of brain waves, blood oxygen level, heart rate, breathing and eye movement is collected, traditionally through a wired system, and used to analyse the sleep stages. Proposed MBAN architectures for monitoring sleep disorders can reduce the complexity of wired monitoring~\cite{Khan:2018}. Haoyu et al.~\cite{Haoyu:2019} have proposed an automatic sleep apnea diagnostic system, based on pulse oximetry and heart rate.

\textbf{Assessing fatigue and athletic readiness:} Stressful situations and the rush of adrenaline might mask the exhaustion and fatigue of the body. A combination of lactic acid and motion sensors can be used to assess physical readiness and bodily fatigue~\cite{Khan:2018}. Such a system can also be useful in the training phase of athletes, where the collected data can be used to optimize workout intervals and rehabilitation times.

\subsubsection{Remote control}

MBANs offer a wide variety of remote monitoring and telemedicine applications. These remote capabilities allow for exciting concepts, such as \textit{Ambient Assisted Living} (AAL), where the monitored data is stored on a back-end medical network~\cite{Movassaghi:2014} and care decisions are made based on that data. This helps the elderly and people in need of care to prolong the home-care period, delaying the need for treatments in medical facilities. The real-time monitoring feature of MBANs can also be used to track recovery processes and remotely administer medications if needed.

%% file: 03_MBAN_security_concerns.tex
\section{MBAN security attributes \& threats}
\label{chap:mban_security_requirements_and_threat_landscape}

Having presented the very basics needed to understand MBAN concepts and their inherent challenges, the security and privacy attributes will be discussed next. Given that current MBAN applications are merely conceptual works or still in a research state, security has to be treated with special care, as there are still many unknowns. In this context, the term security describes the protection of gathered data, whether in transit, in use or at rest. Privacy, on the other hand, refers to controlling the usage and collection of said data. When looking at the current research and already available commercial products, it seems that there is a clear focus on functionality and usability. However, given the nature of data handled by MBANs, security and privacy must be given equal emphasis.

\ifnum\hltext>0
\hl{It is important to note that in this section we will provide a holistic view of MBAN security encompassing all the OSI layers even though IEEE 802.15.6 deals with only PHY and MAC layers. We feel that such a treatment is important since we cannot look at security in a standalone manner due to the interplay that exists between multiple layers. In Section~{\ref{sec:IEEE_802.15.6_security}}, however, we will zoom in and stay inside the purview of the standard.}
\else
It is important to note that in this section we will provide a holistic view of MBAN security encompassing all the OSI layers even though IEEE 802.15.6 deals with only PHY and MAC layers. We feel that such a treatment is important since we cannot look at security in a standalone manner due to the interplay that exists between multiple layers. In Section~\ref{sec:IEEE_802.15.6_security}, however, we will zoom in and stay inside the purview of the standard.
\fi

\subsection{Security attributes}
\label{sec:security_requirements}

The rising numbers of IoT devices in combination with the often extremely sensitive information pose an attractive \textit{attack surface} for potential adversaries. When looking at MBANs, the so-called \textit{Personal Health Information} (PHI) transported has the highest level of sensitivity (according to the ISO/IEC 29100 standard), thus requiring not only increasing level of technical controls to secure this data but also more trust from the user. If potential users of this technology cannot be absolutely sure that only authorized parties can access their PHI, general adoption and acceptance of this ecosystem will be low. There are already several methodologies and approaches on how to tackle security and privacy concerns in not only medical applications but IoT devices in general. In this section, we build upon the necessary \textit{security attributes} for traditional IMDs~\cite{siddiqi2018attack,siddiqi2020imdfence,siddiqi2021thesis} and expand them accordingly for MBANs.

\textbf{Confidentiality (S1):} Data confidentiality can be compared to privacy, meaning that certain information should only be available to certain people. It prevents unauthorized users from accessing the data while at the same time ensuring access for legitimate users. PHI is the most sensitive kind of information available. If leaked, it can have a variety of social and economic repercussions for the victim. A common practice is to categorize the data by its level of sensitivity and implement more or less stringent security controls accordingly. The most common control implemented to ensure confidentiality of data is encryption.

\textbf{Integrity (S2):} means that the processed data is stored and transported as intended and any modification to that data can be detected. By protecting data from alteration, trustworthiness and accuracy is ensured. Falsified information can have serious consequences. For example, if falsified information is sent to medical professionals, they might wrongfully decide to deliver a treatment.

\textbf{Availability (S3):} Given the criticality of the data processed by an MBAN, it must be accessible to authorized users at all times. \textit{Denial of Service} (DoS) attacks~\cite{siddiqi2019towards,siddiqi2020zero} (e.g., jamming, flooding, battery depletion etc.) can make data unavailable, which can lead to failure in delivering treatment, potentially causing a life threatening situation for the patient.

\textbf{Authentication (S4):} Nodes within an MBAN must have the ability to identify the sender and verify if the data received is from a trusted source. Authentication is usually ensured by employing Message Authentication Codes (MAC) and digital signatures, which also ensure data integrity (S2).

\textbf{Authorization (S5):} After successfully authenticating the identity of the sender, it has to be decided which actions can be performed and which resources can be accessed. Authorization is needed to decide who can access and manipulate data in the medical database and within the MBAN. Clusters of sensors, for instance, sometimes need to retrieve different data than the rest of the nodes in the network, giving them a different authorization level~\cite{Doss:2016}.

\textbf{Accountability and Non-repudiation (S6):} Given the sensitivity of the processed data, it has to be visible who has access and who can manipulate it. Data users (e.g., patients, medical professionals etc.) need to be held accountable in case they abuse their privilege to carry out unauthorized actions~\cite{Li:2010}. Additionally, they must not be able to refute the fact that they were the ones accessing or manipulating the data.

\textbf{Data freshness (S7):} Data freshness ensures that old data is not recycled and that data frames belong to the current session~\cite{Al:2017}. When certain identifiers of transmitted data are not unique and get reused, there is a danger that someone records said data and replays it at a later point in time.

\textbf{Dependability (S8):} Dependability is a critical concern in MBANs as it guarantees retrievability of data even in case of failures or a malicious node modification. A failure to retrieve the correct data can interfere with the ability to deliver correct treatment to the patient. One possibility to address this issue is by using error-correcting-code techniques~\cite{Li:2009}. Even though this is a pressing issue in designing secure and reliable MBANs, it has not received much attention yet~\cite{Khan:2013}.

\textbf{Flexibility (S9):} The MBAN should be able to change access rights depending on the circumstances and the network's environment, i.e., the network must be context-aware. On one hand, the network should be able to adapt to changing access points and control units (e.g., changing topology when the main CCU, such as a cellphone, is not in reach). On the other hand, in emergency or other situations, where a new individual needs to make changes to the data (e.g., paramedics), proper access needs to be granted and access control lists need to be updated. This is very challenging, since malicious actors could impersonate a new doctor and grant themselves access to the network.

\textbf{Robustness (S10):} This means that the MBAN has to be resilient against attacks over all layers of security. After an attack the network should -- at a minimum -- guarantee continuation of function and operability. In contrast to dependability (S8), robustness is focused on the system's behavior rather than its functionality.

\textbf{Secure-key management (S11):} Most MBAN systems rely on some kind of secret key, which is used for encryption, authentication and integrity checks. The main challenge is the distribution of such keys to each node. Moreover, they must be generated by using truly random numbers to alleviate the risk of key replication and be stored in a way inaccessible to any person other than the MBAN user. A prominent approach in MBANs is to use physiological signals to generate such keys. When using asymmetric cryptography, typically a \textit{Public Key Infrastructure} (PKI) is employed, which embeds a public key in a certificate. However, traditional certificates use a lot of memory space, making them a sub-optimal choice for most MBAN nodes. Many alternatives to classical PKIs have been suggested, like TinyPK, $\mu$PKI and L-PKI, where the latter is considered to be most suitable for WBANs in general~\cite{Doss:2016}. Furthermore, certificate-less solutions have also been proposed~\cite{Liu:2013, Shen:2018}.

\subsection{Privacy attributes}
\label{sec:privacy_requirements}

\begin{figure*}[!t]
\centering
  \includegraphics[scale=.8]{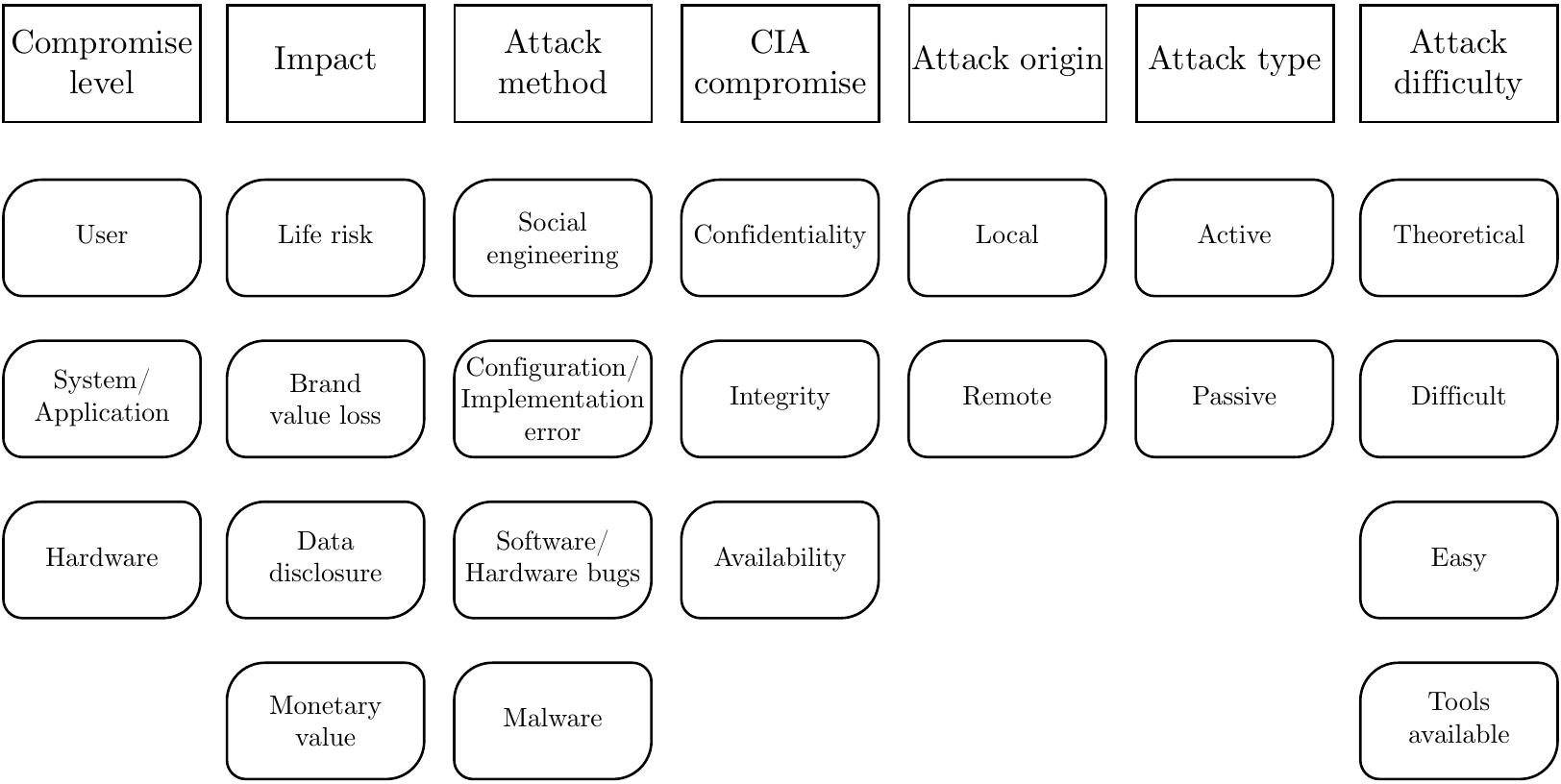}
  \caption{A taxonomy of possible attacks on an MBAN}
  \label{fig:mban_attack_taxonomy}
\end{figure*}

As mentioned previously, privacy of patient-related data must be guaranteed at all times to increase user trust and acceptance. Before deployment of MBAN applications, some important issues must be tackled~\cite{Gupta:2020}; for instance, how data is stored? Who has access to the patient's medical records? How is data handled in emergencies? etc. Several regulations are in place to ensure privacy of medical data. The \textit{General Data Protection Regulation} (GDPR) in Europe and the \textit{Health Insurance Portability and Accountability Act} (HIPAA) in the USA provide a solid framework to correctly handle healthcare-related information, including both civil and criminal consequences~\cite{Al:2017}. In addition to the existing frameworks, MBAN applications that handle \textit{Personal Identifiable Information} (PII) have to comply with the following principles, as proposed in~\cite{hatzivasilis2019review}:

\begin{itemize}
    \item[--] The volume of raw data collection and the overall data volume requested by applications must be minimized, e.g., by lowering sampling rate, bits per sample, recording duration etc.
    \item[--] Individual users should not be identified, unless there is an explicit need.
    \item[--] Collected information should be stored in a confined manner and for as short duration as possible.
    \item[--] As much data as possible should be anonymized, minimizing the amount of exposed PII.
    \item[--] Data should be low in granularity and encrypted when at rest.
\end{itemize}

However, it is important to note that being compliant to privacy regulations does not imply being secure. Data stored in MBANs may be leaked by physically compromising the system or independent nodes. Therefore, the security and privacy attributes need to have a symbiotic relationship.

\subsection{Security tradeoffs}
\label{sec:security_tradeoffs}

Although security is a crucial aspect in designing any kind of MBAN, it cannot be the main concern of the technology. Security needs to be an assisting metric that supports functionality, usability and safety. After all, a technology is not used because it is secure, but because it can add value to existing processes or even revolutionize the way things are done. That is why there are several tradeoffs between security and characteristics that make systems more vulnerable.

\subsubsection{Security vs. Usability}
\label{subsec:security_versus_usability}

For MBAN applications it is of utmost importance that the margin of user mistakes is as low as possible. User interactions should be foolproof as most of the times the input is not provided by experts. For instance, when implementing node pairing mechanisms
\ifnum\hltext>0
\hl{(to e.g., ensure S4 and S11)}
\else
(to e.g., ensure S4 and S11)
\fi
in MBANs, the bootstrap has to involve some manual interaction. More specifically, directly applying device pairing requires $O(n^2)$ human interactions~\cite{Li:2010}. Omitting this human component can degrade the security of the whole system.

\subsubsection{Security vs. Accessibility}
\label{subsec:security_versus_accessibility}

If a patient, who is equipped with an MBAN, is unconscious or incapacitated, they cannot help the paramedics to access the MBAN (e.g., by unlocking the CCU).
In this scenario, security is actually detrimental to the well being of the patient. Therefore, controls need to be implemented to grant accessibility whenever needed, while ensuring that in doing so the attack surface is not broadened for adversaries
\ifnum\hltext>0
\hl{(see S9 under Section~{\ref{sec:security_requirements}})}.
\else
(see S9 under Section~\ref{sec:security_requirements}).
\fi

\subsubsection{Security vs. Resource Limitations}
\label{subsec:security_versus_resource_limitations}

As mentioned previously, MBAN nodes have very limited resources that need to be distributed between value adding functionality, maintenance functions and security controls. A strong security control, i.e., the one that tries to implement the attributes of sections~\ref{sec:security_requirements} and~\ref{sec:privacy_requirements}
\ifnum\hltext>0
\hl{(mainly S1, S2, S4, S6 and S11)},
\else
(mainly S1, S2, S4, S6 and S11),
\fi
also needs correspondingly high amounts of resources. This issue is especially pressing in MBAN applications, where nodes are very often not rechargeable and need to be replaced when out of battery~\cite{zheng2016ideas}.

\subsection{Threat landscape}
\label{sec:threat_landscape}

In order to design secure MBANs, it is necessary to exactly know the attack surface offered to an adversary. Since no system will ever be perfectly secure and new vulnerabilities will always be found, attack methods are constantly evolving. As one can imagine, there are several possibilities for categorizing the plethora of existing attacks. Alsubaei et al.~\cite{alsubaei2017security} introduce a taxonomy of attack types on IoMT devices, which is shown in Figure~\ref{fig:mban_attack_taxonomy}.

\begin{figure}[!t]
\centering
  \includegraphics[scale=.5]{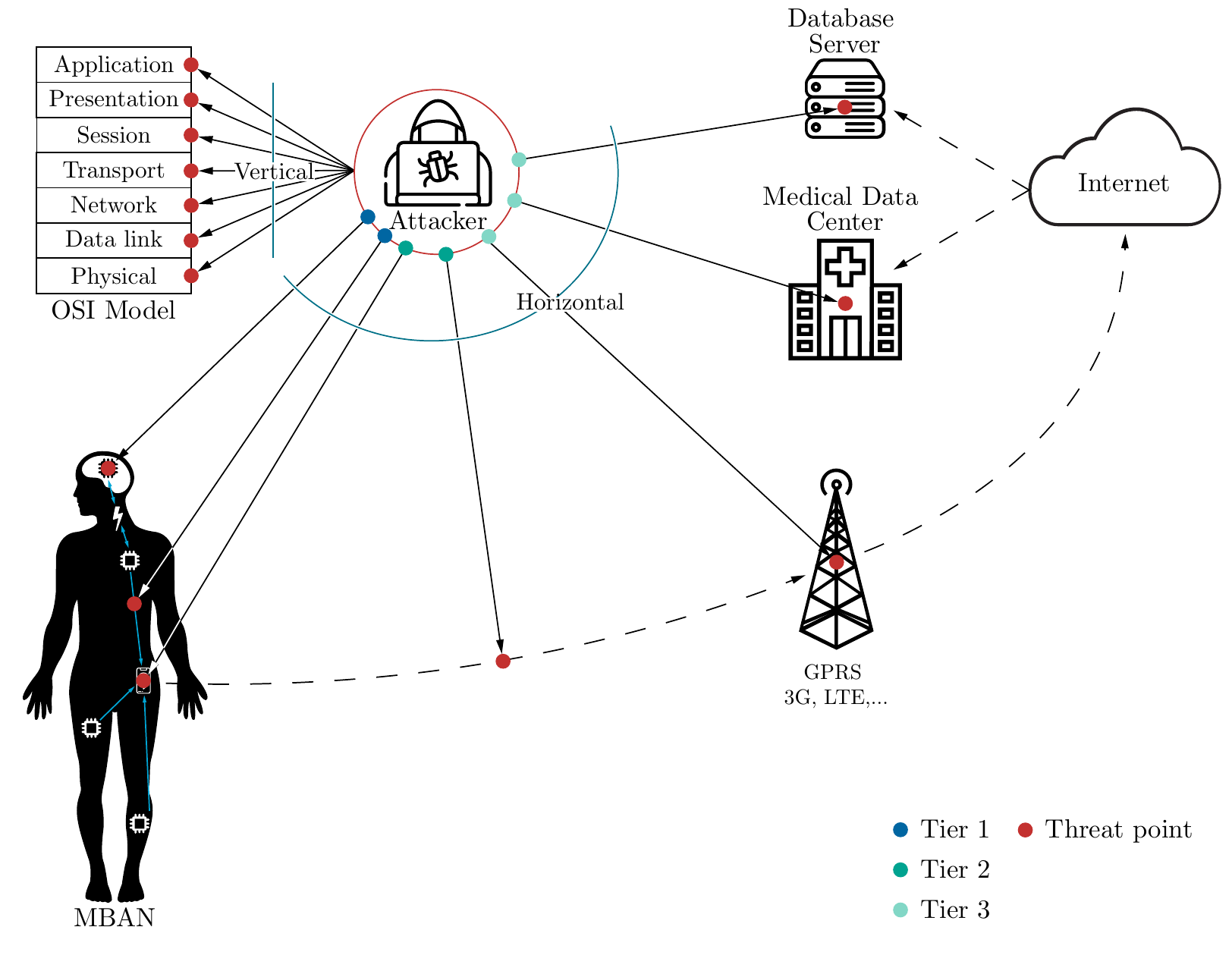}
  \caption{An overview of the entry points into an MBAN offered to an attacker}
  \label{fig:wban_threat_points}
\end{figure}

In MBANs, the attack surface can be split into horizontal and vertical panes, giving the attacker a two-dimensional perspective. There are three distinct horizontal entry points, which coincide with the three tiers introduced in Section~\ref{sec:WBAN_architecture}. Each such point of entry is accompanied by a vertical pane, which consists of the seven-layer \textit{OSI} model~\cite{Doss:2016}. Figure~\ref{fig:wban_threat_points} shows the possibilities an attacker has to enter the network. Using these entry points the exact attack surface for an MBAN can be discussed. These attacks appear at the hardware, software and network-protocol stage of each element in the corresponding tier.

\begin{table*}[!t]
    \centering
    \caption{A complete overview of demonstrated attacks on nodes typically used in MBAN applications}
    \label{tab:mban_demonstrated_attacks}
    \resizebox{\textwidth}{!}{%
    \begin{tabular}{p{1.5cm}p{1.4cm}lp{1.8cm}lp{4cm}p{4cm}}
        \hline
        MBAN node & Type & Functionality & Attack type & Layer & Vulnerability & Exploit \\ \hline
        Animas OneTouch Ping Insulin Pump~\cite{rapid7insulinpump:2016} &
        Semi-invasive &
        Actuator & Eavesdropping & Transport &
        Packets between the remote node and pump are sent as clear text. &
        An attacker can eavesdrop on the traffic transmitted by the device and capture packets containing blood glucose and insulin dosage data. \\
        &
        &
        &
        Impersonation & Transport &
        The CRC32 key used for pairing and authentication is statically used, and transmitted in the clear. &
        Attackers can sniff the CRC32 key and impersonate the remote node. This would allow them to remotely administer doses of insulin. \\
        &
        &
        &
        Replay & Network &
        Communication between the pump and the remote node do not have any kind of replay protection (e.g., sequence numbers, packet identifiers, timestamps etc.). &
        An attacker can capture commands sent to the pump and replay them at a later point in time without any specific knowledge about the packet structure. \\ \hline
        Older-generation ICDs~\cite{halperin2008pacemakers} &
        Implantable & Hybrid & DoS & Data link &
        The short-range communication protocol is sent over the air as clear text and can be reverse engineered. &
        The wake-up protocol of the ICD can be exploited to continuously activate the RF module and thereby draining the battery of the device. \\ \hline
        Newer-generation ICDs~\cite{marin2016security} &
        Implantable & Hybrid & Replay/ Spoofing & Transport &
        By intercepting the long-range communication between the ICD and the programming wand, messages can be reverse engineered due to the device relying on \textit{security through obscurity}. Intercepted messages always have the same header. &
        Intercepted messages can be eavesdropped by an attacker wearing a backpack with the right equipment. This data can then be replayed at a later point in time while being relatively close to the patient (e.g., in public transport). \\
        &
        &
        &
        Eavesdropping & Transport &
        Sensitive patient data transmitted over the air is \textit{obfuscated} by using a static Linear-Feedback-Shift-Register (LFSR) sequence. &
        Attackers can passively eavesdrop the channel during an ongoing transmission and possibly gather private information about the patient. Eavesdropped information can be used to track, locate and identify patients. \\
        &
        &
        &
        Impersonation/ DoS &
        Data link &
        The device does not immediately go to \textit{sleep} mode after finishing communication, but to \textit{standby} mode for five minutes. While in standby, the device can be activated by sending a message, which is always the same. &
        It is possible for an attacker to impersonate the device programmer and repeatedly send \textit{wake-up} calls to drain the battery or block legitimate traffic to compromise patient safety. \\ \hline
        FitBit fitness tracker~\cite{classen2018anatomy} &
        Wearable & Sensor &
        Man-in-the-Middle &
        Application &
        Login credentials are sent in plain text and just secured by HTTPS without MITM protection. &
        Through modification of the smartphone app, attackers can associate trackers to another FitBit account and steal trackers. \\
        &
        &
        &
        Firmware customization &
        Application &
        The BLE connection has Generic Attribute Profile (GATT) enabled, making it possible to remotely flash custom firmware on the device. &
        Custom firmware can override security protocols, making it possible to leak sensitive data to an attacker. \\ \hline
        Hospira Symbiq infusion system~\cite{us2015cybersecurity} &
        Semi-invasive &
        Actuator &
        Tampering/ Modification &
        Application &
        Pumps do not check incoming updates for authenticity. Corrupted libraries can be uploaded through the hospital network. &
        An attacker can transmit malicious commands to the infusion system, potentially directing the pump to perform unanticipated actions. \\ \hline
        Hermes medical shoe~\cite{yan2014semantic} &
        Wearable &
        Sensor &
        Tampering/ Modification &
        Transport &
        The pressure-sensor data and the time between transmissions can be altered with sufficient access to the platform. &
        An attacker can tamper with the original data to alter the diagnostic decision-making process. \\ \hline
        Drop sensor infusion pumps~\cite{park2016ain} &
        Semi-invasive &
        Actuator &
        Sensor spoofing/ DoS &
        Physical &
        Drop sensors are susceptible to signal injection of a spoofing signal using the same physical quantity. Alarm systems can be bypassed by using the right signal patterns. &
        By injecting an external high power signal into the drop sensor, the output can go into saturation, causing the drop counting mechanism to fail. \\ \hline
    \end{tabular}%
    }
\end{table*}

The connection between the gateway and the Internet is a popular target for attackers, as it is accessible from the outside world. However, this link is usually protected by state-of-the-art security protocols, like SSL/TLS and IPSec. As these protocols are not MBAN specific, they will be treated as a black box. Similarly, an attacker could attempt to infiltrate the data storage or cloud servers directly and steal sensitive information of multiple patients. But, as in the previous attack vector, this area of the network should be secured by adequate access-control mechanisms, robust encryption protocols and proper authentication.

Once the correct entry point is chosen, two kinds of actions can be initiated, namely, active and passive attacks. In passive attacks, data is only received and not written to the data stream, thus making them less intrusive. Active attacks on the other hand read and write to the data stream, possibly causing data corruption or Denial of Service. From an attacker's point of view, both attack types have their advantages. Results obtained by active attacks might be more valuable and impactful, while passive attacks are often very stealthy and hard to detect.

Table~\ref{tab:mban_demonstrated_attacks} gives an up-to-date overview of the demonstrated attacks that have been conducted on Tier 1 MBAN nodes (see Figure~\ref{fig:detailed_architecture_of_WBAN}). As can be seen, the attack surface that is offered by these MBAN applications includes vulnerabilities at all the layers of the OSI model.

%% file: 04_The_IEEE_802.15.6_standard.tex
\section{IEEE-802.15.6 security}
\label{sec:IEEE_802.15.6_security}

\begin{figure*}[!t]
\centering
  \includegraphics[scale=.8]{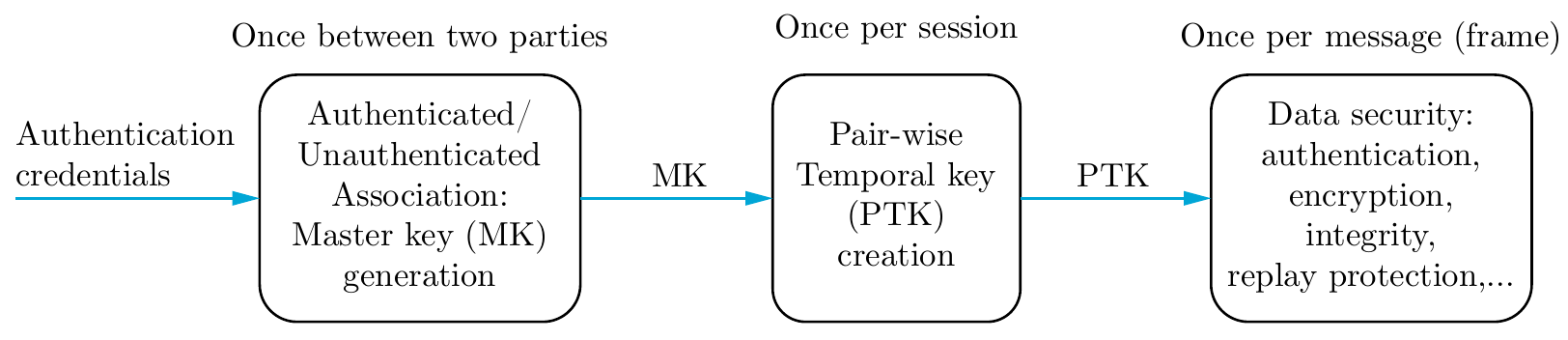}
  \caption{Security hierarchy as specified in IEEE 802.15.6~\cite{ieee_standard}}
  \label{fig:IEEE_802_15_6_security}
\end{figure*}

\begin{figure*}[!t]
\centering
  \includegraphics[scale=.9]{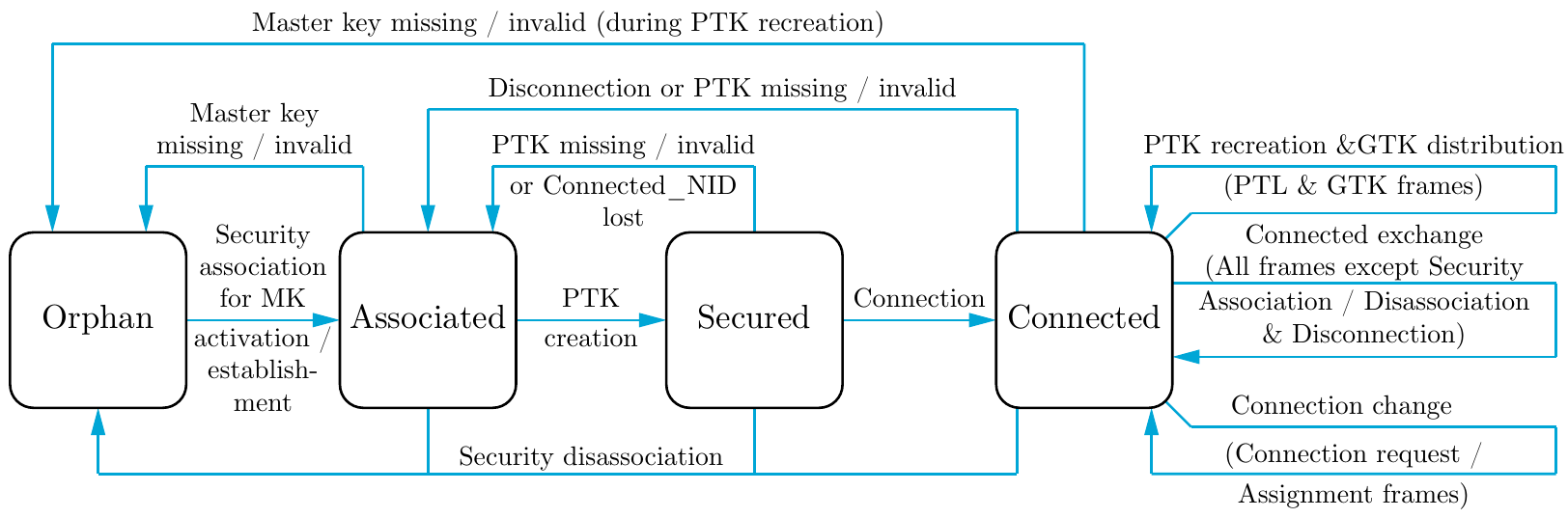}
  \caption{IEEE 802.15.6 MAC layer security state diagram for secured communication~\cite{ieee_standard}}
  \label{fig:IEEE_802_15_6_secure_communication}
\end{figure*}

In this section, we will focus on the security aspects of the IEEE 802.15.6 standard, which are important to consider given the MBAN threat landscape discussed in Section~\ref{chap:mban_security_requirements_and_threat_landscape}.
IEEE 802.15.6 offers three distinct security levels, from which hubs (nodes managing the network's traffic) have to choose one during security association:

\textbf{Level 0 - Unsecured communication:} Messages are transmitted in unprotected data frames. The lack of privacy protection and security mechanisms leave the network wide open to a variety of attacks.

\textbf{Level 1 - Authentication without Encryption:} Security measures are implemented to guarantee that only authenticated entities can receive, transmit or manipulate data frames. Message confidentiality and privacy are however not ensured, as the data is not encrypted.

\textbf{Level 2 - Authentication and Encryption:} This level combines the implemented authentication mechanisms with state-of-the-art encryption algorithms to protect the confidentiality and privacy of transmitted data. Therefore, this method offers the most extensive protection against attacks~\cite{Ullah:2013}.

The \textit{Security Suite Selector} (SSS) field in the security association frame is used to help in choosing the security level, the security-association protocol and the employed cipher function. It can also indicate whether control frames need to be authenticated.

\subsection{Security hierarchy and secured communication}
\label{subsec:security_hierarchy_and_secured_communication}

The standard offers a clear guideline and implementation of how and when cryptographic keys are established and activated. Figure~\ref{fig:IEEE_802_15_6_security} shows the basic security hierarchy of IEEE 802.15.6, which is used to identify nodes and hubs. For this purpose, either a pre-shared or a newly established master key (MK) is activated to open secure communications. In the case of unicast communication, a pairwise temporal key (PTK) is created and shared once per session. For multicast communication, a group temporal key (GTK) is created and subsequently shared with the relevant group using the unicast method~\cite{kwak2010overview, Ullah:2013}.

Before establishing a secure communication to exchange data, a node-hub pair passes certain stages at the MAC level. Figure~\ref{fig:IEEE_802_15_6_secure_communication} shows the state diagram specified in the standard.
As can be seen, when establishing secured communication, MBAN nodes can be in four distinct states: \textit{orphan}, \textit{associated}, \textit{secured} and \textit{connected}.

\subsection{Security-association and -disassociation protocols}
\label{subsec:security_association_and_disassociation_protocols}

In order to establish a secure connection between nodes in the network, the standard offers five distinct \textit{Authenticated Key Exchange} (AKE) and \textit{Password Authenticated Key Exchange} (PAKE) protocols for association: \textit{(I) pre-shared MK}, \textit{(II) unauthenticated}, \textit{(III) public-key hidden}, \textit{(IV) password authenticated association} and \textit{(V) display authenticated}. Furthermore, the standard employs one protocol for \textit{PTK creation / GTK distribution (VI)} and one for the \textit{disassociation procedure (VII)}. Whereby, association and disassociation describe the mechanisms for exchanging and erasing master and pair-wise temporal keys between a node and a hub, respectively. Protocols typically consist of a three-phase handshake, namely, request, response and activate (or erase). In this context, we call the party sending the first frame the Initiator \textit{I} and its counterpart the Responder \textit{R}.

The above protocols, except for pre-shared MK, implement the Elliptic-curve Diffie--Hellman (ECDH) key exchange mechanism using the P‐256 curve from the FIPS Pub 186‐3 secure hash standard.

If transported in a secure mode, message frames are encrypted using AES-128 in counter with cipher block chaining (CCM) mode, in which a 13-octet nonce, containing both high and low order sequence numbers, is required for each session to synchronise frames in order to mitigate replay attacks and guarantee data freshness~\cite{Ullah:2013}. Alternatively, the standard also offers the option to use the slower 128-bit Camellia cipher. However, this cipher does not offer any obvious advantages over its counterpart and it is far less tested and established.

\subsection{Vulnerabilities and weaknesses}
\label{subsec:security_vulnerabilities_and_weaknesses}

For any novel technology, establishing a reliable and secure framework around it needs to be an iterative process as new zero-day vulnerabilities will most likely be found. This is even true for protocols that have undergone rigorous testing and validation.
The IEEE 802.15.6 standard is no exception to this. Though it offers an immense potential for future MBAN applications, there are still some key weaknesses that need to be addressed.

The standard is based on an WBAN architecture in which a hub is the central coordinator of the network: All nodes are directly connected to it, creating a star topology. However, if the hub is not able to communicate with the nodes, the network stops functioning as a whole. This offers an attacker the opportunity to exhaust the hub's resources by sending it a large number of invalid frames. Moreover, since the hub needs to be equipped with superior computing, memory and energy resources, it is most likely not implanted within the body, making it even more accessible to the outside world. Thus, physical theft or damage to the hub can have the same Denial-of-Service effect.
In addition, several severe vulnerabilities in the standard's association protocols were found by Toorani~\cite{toorani2015vulnerabilities, toorani2016security}.

%% file: 05_Exploring_future_MBAN_application_scenarios.tex
\section{IEEE-802.15.6 security analysis \& assessment}
\label{chap:assesing_and_analysing_the_standard}

To assess the IEEE 802.15.6 standard's security weaknesses and shortcomings across a large number of dimensions, we adopt a structured analysis method. We will begin by explaining our analysis and assessment strategy, and will subsequently apply it to the standard. We will conclude the section by offering specific recommendations with the aim of improving the standard in future iterations.

\subsection{Introducing the assessment methodology}
\label{sec:introducing_the_assessment_procedure}

Our assessment relies on the realization that each security attribute (Sx) defined in section~\ref{sec:security_requirements}, will have additional repercussions on the targeted design, i.e., on the supplementary physical attributes (Px) and the eligible device classes (Dx), i.e., the node types classified by implementation from Section~\ref{sec:node_types}: invasive, semi-invasive, wearable and ambient. The permutations of the values of these \textit{security attributes (Sx)} and \textit{security-related attributes (Px \& Dx)} -- cumulatively termed henceforth as security(-related) attributes -- create a large design space (of possible MBAN implementations) that needs to be traversed in order to guarantee the completeness of the analysis. However, an exhaustive traversal even when using a limited number of attributes quickly becomes intractable. Even by pruning out large design-space regions not corresponding to sensical MBAN system instances, the approach is highly impractical.

To resolve this problem, we adopt a heuristic approach: we first identify and describe realistic use cases whose \textit{design specifications} cumulatively cover the feasible part of the design space. We then validate our consolidated security(-related) attributes against these use cases to find the standard's shortcomings. The heuristic aspect of this approach lies in the assumption that the set of selected use cases \textit{completely covers} the whole field of MBAN applications. It also lies in the fact that, by adding more use cases or more security(-related) attributes in the future, the coverage of the design space can be improved at will, by paying for an exponentially growing validation effort.

In this work, we have compiled an already extensive list of Px and their subcategories, shown in Table~\ref{tab:physical-attributes}: computational capability Cx, memory capacity Mx, energy source Ex and network topology Tx. Obviously, defining \textit{absolute} measures for the computing, storage or other capabilities of an MBAN node is extremely difficult and, even if achieved, it would have to be constantly redefined subject to continuous technological advance. Thus, the quantifications defined for Px,  e.g., \textit{full} vs. \textit{moderate} computing, are arbitrary and intended to conceptually handle different \textit{classes} of MBAN nodes and whole applications. Without loss of generality, the same methodology can be extended to account for, e.g., \textit{five} different levels of computing capability, instead of only three, or even for precise numerical ranges if so desired, albeit at the cost of a finer-grain, slower analysis.

\begin{table}[!t]
    \centering
    \caption{An overview of the security-related physical attributes (Px) used in the security-assessment methodology}
    \label{tab:physical-attributes}
    \footnotesize
    \begin{tabular}{llc}
        \hline
        \multicolumn{1}{l}{Physical attribute} & \multicolumn{1}{l}{Attribute value} & \multicolumn{1}{c}{Notation}  \\ \hline
        \multirow{ 3}{*}{Energy source}        & Passive   & E1         \\
               & Non-rechargeable battery   & E2         \\
               & Rechargeable battery   & E3         \\ \hline
        \multirow{ 3}{*}{Memory capacity}        & No/light memory   & M1         \\
               & Moderate memory   & M2         \\
               & Full memory   & M3         \\ \hline
        \multirow{ 3}{*}{Computational capability}        & No/light computation   & C1         \\
               & Moderate computation   & C2         \\
               & Full computation   & C3         \\ \hline
        \multirow{ 3}{*}{Network topology}        & Star   & T1         \\
               & Tree   & T2         \\
               & Peer to peer   & T3         \\
        \hline
    \end{tabular}
\end{table}

\begin{figure}[!t]
\centering
  \includegraphics[trim={7cm 2cm 0cm 2cm},clip,scale=0.42]{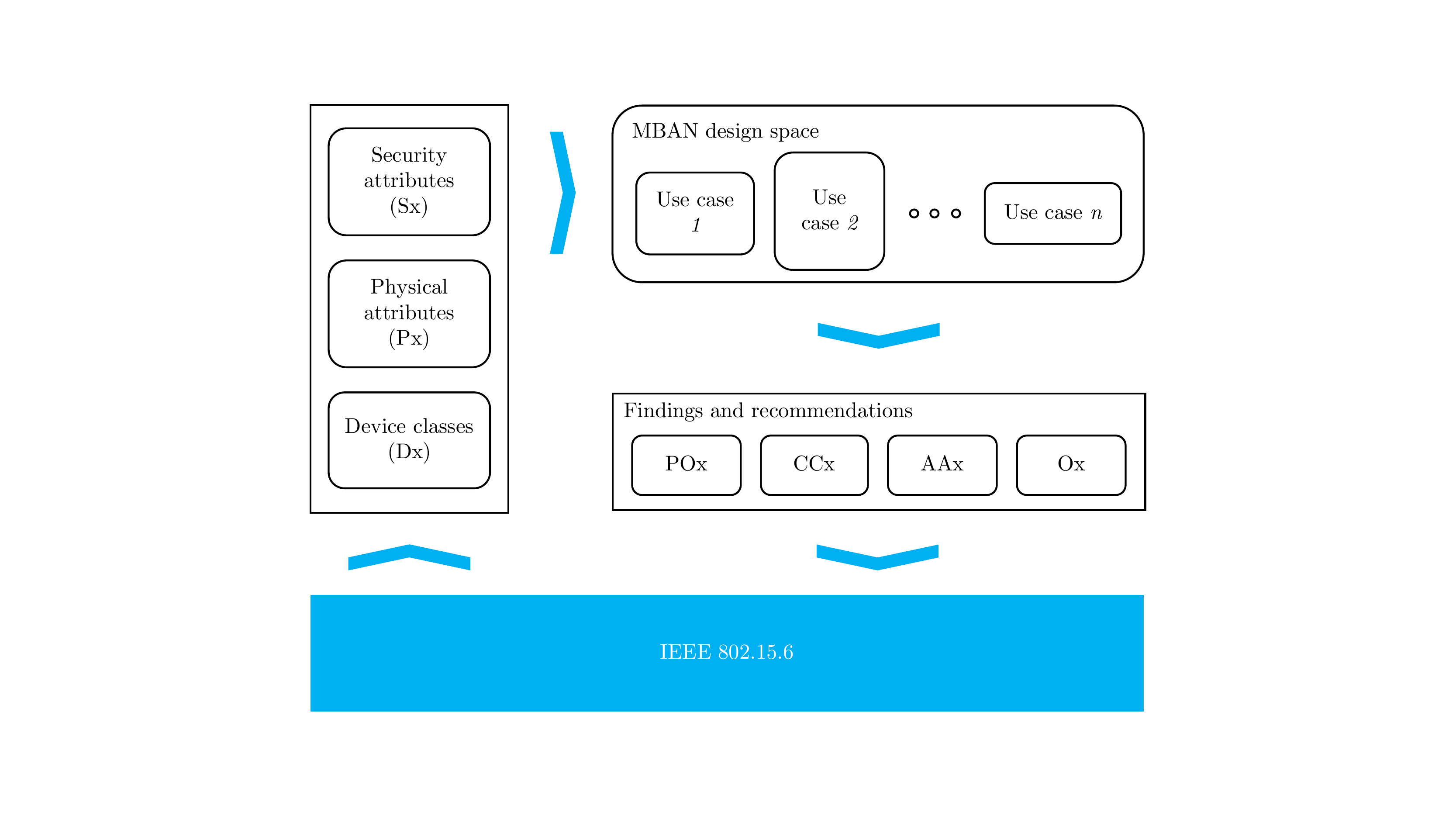}
  \caption{An overview of the different attributes used in the security-assessment methodology and their relationship with the use-case specifications (Ux.x) and recommendations to the standard}
  \label{fig:relationship_requirements}
\end{figure}

Figure~\ref{fig:relationship_requirements} illustrates the relationship between the security(-related) attributes (i.e., Sx, Px and Dx) and the use-case specifications. As alluded to in the figure, the next step is to design rational and useful use cases covering the entire spectrum of defined attributes. Therefore, a two-dimensional matrix is created, as will be shown in Section~\ref{sec:introducing_the_hypothetical_scenarios} (Table~\ref{tab:master_matrix}). This matrix connects the range of attributes and the specific components (i.e., nodes) of each use case. Only if the use cases collectively require the fulfillment of each security(-related) attribute, they can be considered as collectively exhaustive.

After the design of the use cases is complete, for each use case a number of use-case specifications (Ux.x) are formulated. These specifications are connected to the previous attributes (Sx and Px) by using a similar matrix that will be discussed in Section~\ref{sec:introducing_the_hypothetical_scenarios} (Table~\ref{tab:application_specific_matrix}). This makes it possible to validate that the newly formulated use-case specifications are covering the entire range of relevant dimensions, guaranteeing a comprehensive foundation for the subsequent in-depth analysis of the standard.
Whether the list of Ux.x is complete or not does not matter in the context of this analysis as long as the Ux.x of each use case collectively cover Sx and Px.

\ifnum\hltext>0
\hl{The recommendations that appear as a consequence of this analysis are clustered into the following categories:}
\else
The recommendations that appear as a consequence of this analysis are clustered into the following categories:
\fi

\begin{itemize}
    \item Physical and organizational
    \ifnum\hltext>0
    \hl{(POx)}
    \else
    (POx)
    \fi
    \item Cryptography, confidentiality and integrity
    \ifnum\hltext>0
    \hl{(CCx)}
    \else
    (CCx)
    \fi
    \item Authentication and authorization
    \ifnum\hltext>0
    \hl{(AAx)}
    \else
    (AAx)
    \fi
    \item Other
    \ifnum\hltext>0
    \hl{(Ox)}
    \else
    (Ox)
    \fi
\end{itemize}

An overview of the relationship between Ux.x and recommendations can be seen in Figure~\ref{fig:relationship_requirements}.

\subsection{Defining representative MBAN use cases}
\label{sec:introducing_the_hypothetical_scenarios}

Since we have already defined all of the necessary attributes (Sx, Px and Dx), specifying representative MBAN use cases is the next step of the assessment process. While it is possible to create completely fictional use cases which cover all of the necessary dimensions, this section concentrates on real-life, actively researched applications. These use cases are ordered by decreasing complexity and are discussed next.

\ifnum\hltext>0
\hl{It should be noted that even though the standard focuses on PHY and MAC layers, we will zoom out and also include the use-case specifications that lie outside these layers. This is based on the holistic approach of Section~{\ref{chap:mban_security_requirements_and_threat_landscape}}.}
\else
It should be noted that even though the standard focuses on PHY and MAC layers, we will zoom out and also include the use-case specifications that lie outside these layers. This is based on the holistic approach of Section~\ref{chap:mban_security_requirements_and_threat_landscape}.
\fi

\subsubsection{Use case 1: Neural Dust}
\label{subsec:scenario_1_neural_dust}

\begin{figure}[!t]
\centering
  \includegraphics[scale=.6]{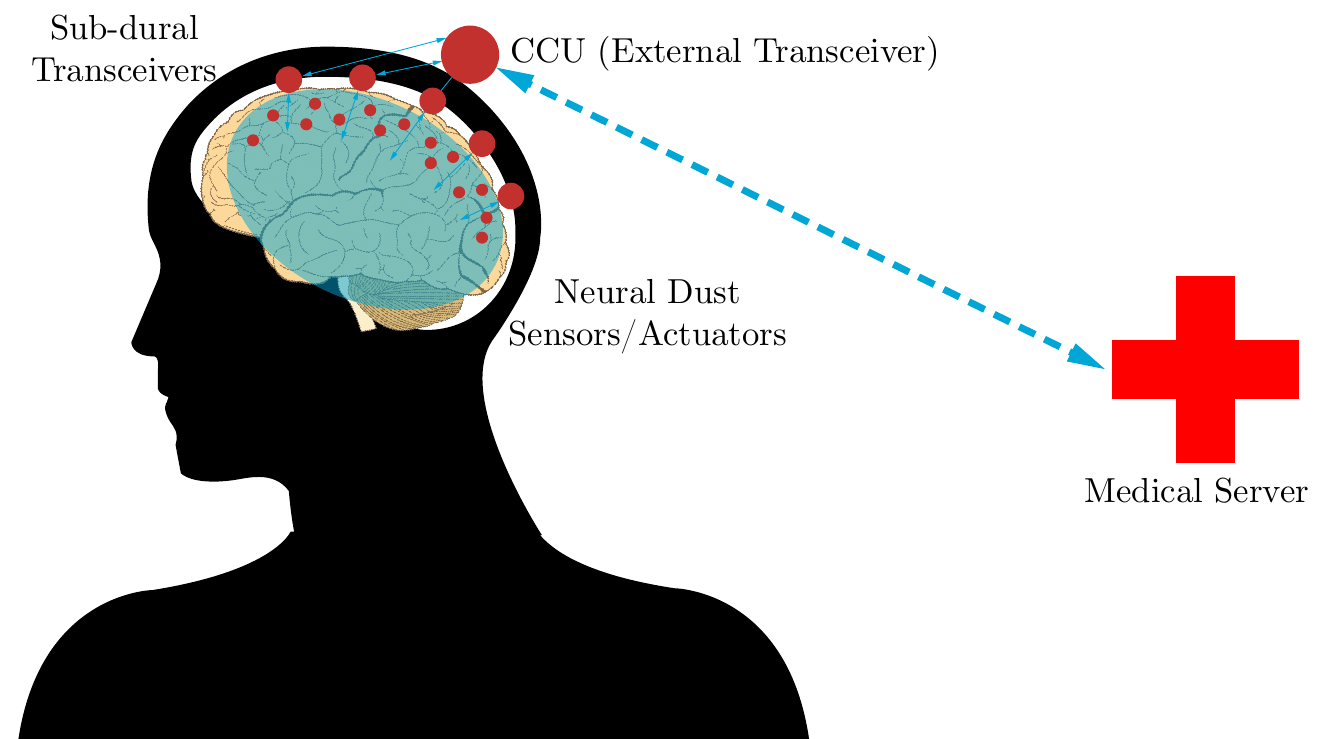}
  \caption{Overview of use case 1: Neural-dust sensors and actuators spread across the brain, communicating with sub-dural transceivers that relay data to and from an external transceiver}
  \label{fig:future_scenario1}
\end{figure}

The concept of smart dust has been around for over 20 years and the initial thought behind it is still relevant. The principal idea is to establish a network of thousands of free-floating, independent, micron-sized sensor and actuator nodes (resembling the \textit{dust}) spread across the brain (neural dust) or in the intestines (body dust) for monitoring and stimulating purposes~\cite{seo2013neural}. The neural-dust concept is revolutionizing the way we think about traditional brain-machine interfaces (BMIs) especially in the context of chronic and long-term treatment. Increased bio-compatibility by massively decreasing the form factor, adding encapsulation, eliminating wired connections and removing the necessity of a battery for implantable nodes make this technology very promising.
There are two main use cases that have been proposed for neural dust applications: (1) Sensory nodes can be implanted in the cortex to chronically record extracellular electrophysiological activities, which can then be sent to a medical server for further diagnostics. (2) Sub-cortically implanted actuator nodes may be used for deep-brain stimulation to treat a variety of diseases, like Alzheimer's and epilepsy. Henceforth, said sensor and actuator nodes implanted into the brain will also be referred to as \textit{dust nodes}. Figure~\ref{fig:future_scenario1} shows the basic topology of the neural dust use case.

The created use case is based on the well-established conceptual research of Seo et al.~\cite{seo2013neural}, which consists of three stages of communication. Multiple transceivers implanted beneath the dura mater communicate with dust nodes implanted into the brain's cortex and sub-cortex via ultrasound signals.
Ultrasound is also used by the sub-dural transceivers to wirelessly transfer power to the dust nodes.
In terms of security, the ultrasound communication between dust nodes and sub-dural transceivers does not employ any encryption algorithms since the dust nodes of the state of the art cannot handle complex computations, including cryptography.
However, MHz-range ultrasound communication was recently validated by Siddiqi et al.~\cite{siddiqi2021securing} to be secure from eavesdropping and message-insertion attacks \textit{unless} the adversary is physically touching the patient.
For this use case, we assume that the cost for an attacker to come in physical contact with the patient and initiate a successful attack is unreasonably high when compared to other possibilities to do harm.

The sole purpose of the sub-dural transceivers is to relay information between the correct dust nodes and a wearable, external receiver acting as the CCU. Though they are only used as relay nodes within the network, they have adequate computational and memory capacities to be able to compute various protocols. Furthermore, it is assumed, that the CCU has the computational means to coordinate the entire network's functionality and to establish a WiFi connection to a nearby router, eliminating the need for a personal device (e.g., mobile phone, smart watch etc.). As sub-dural transceivers and the CCU are separated by the skull, which blocks and attenuates ultrasound waves, an RF channel is used instead for communication. The CCU also utilizes this channel to wirelessly power the sub-dural transceiver. In contrast to the communication between dust nodes and sub-dural transceivers, the communication between sub-dural transceivers and the CCU is fully encrypted.

The use-case specifications are as follows:

\begin{itemize}
    \item \textbf{U1.1:} Consider the passive and highly resource-constrained character of dust nodes and sub-dural transceivers (E1, M1, C1)
    \item \textbf{U1.2:} Support the network's tree (extended two-hop star) topology (T2)
    \item \textbf{U1.3:} Be scalable enough to support a great number of dust nodes (S9)
    \item \textbf{U1.4:} Sufficiently encrypt messages between the CCU and sub-dural transceivers at all times (S1, S2)
    \item \textbf{U1.5:} Support mutual authentication between CCU and sub-dural transceivers (S4)
    \item \textbf{U1.6:} Guarantee that security keys can only be generated and used by legitimate parties (S1, S11)
    \item \textbf{U1.7:} Ensure that the network still functions if dust nodes or a sub-dural transceiver fails or is under a DoS attack (S3, S8, S10)
    \item \textbf{U1.8:} Ensure data frames are protected with non-repeating sequence numbers to mitigate the risk of eavesdropping and replay attacks (S1, S2, S7)
    \item \textbf{U1.9:} Ensure messages can only be delivered to dust nodes via legitimate sub-dural transceivers while ensuring accountability (S5, S6)
\end{itemize}

\subsubsection{Use case 2: Leadless Cardiac Pacemaker}
\label{subsec:scenario_2_leadless_cardiac_pacemaker}

Conventional cardiac pacemakers are amongst the most used IMDs to date.
Usually, they consist of a subcutaneous generator pocket alongside a transvenous lead for cardiac sensing and stimulation. Although this technology is widely established as the standard treatment for symptomatic bradyarrhythmias, complications like pneumothorax, cardiac occlusion, pocket hematoma, lead perforation, fracture and dislodgement can still occur~\cite{poole2010complication, tjong2017permanent, cantillon2017complications}.

Leadless cardiac pacemakers (LCPs) are trying to mitigate the above risks by decreasing the overall size and invasiveness of the components. Currently, there are two clinically available systems, namely the Nanostim Leadless Cardiac Pacemaker from Abbott, and the Micra Transcatheter Pacing System from Medtronic. Both are completely self-contained and capable of providing single-chamber right ventricular pacing, sensing and rate response delivery~\cite{tjong2017permanent}. However, these solutions currently show some functional limitations as there are no capabilities for Cardiac Re-synchronisation Therapy (CRT)~\cite{joury2021leadless}.

\begin{figure}[!t]
\centering
  \includegraphics[scale=.7]{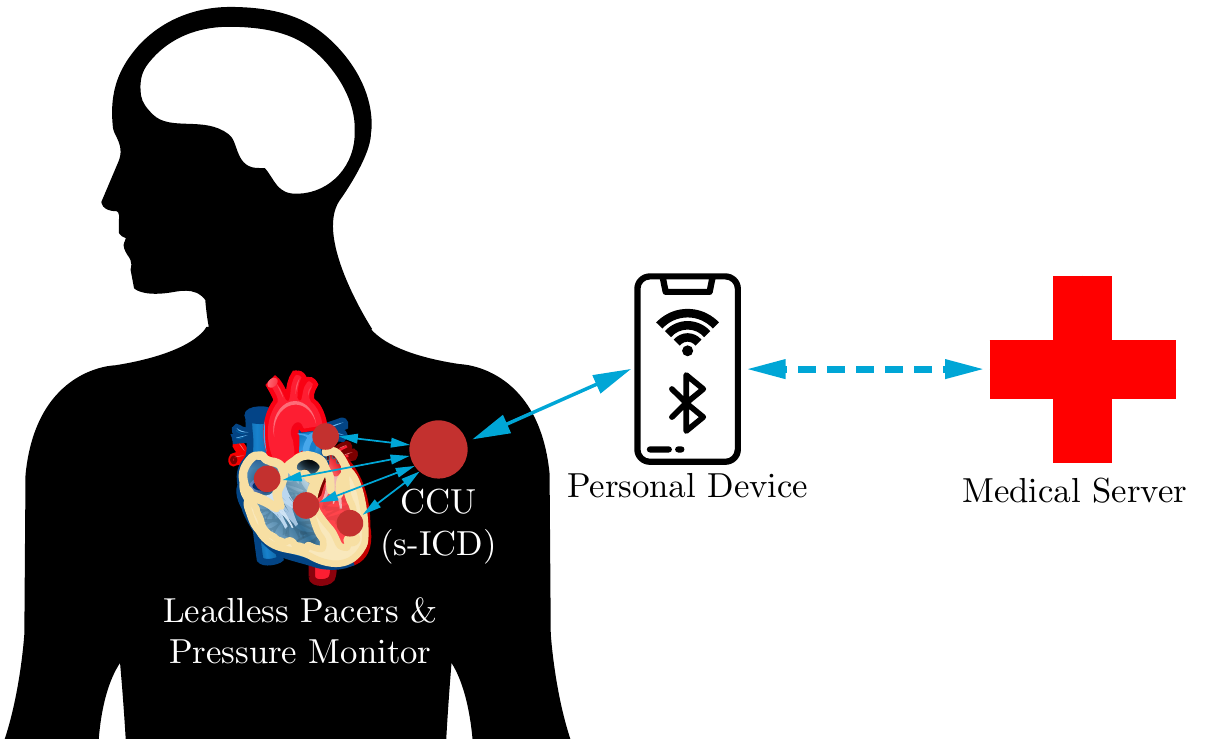}
  \caption{Overview of use case 2: Permanent LCPs combined with an s-ICD that functions as the network's coordinator}
  \label{fig:future_scenario2}
\end{figure}

According to Tjong et al., in the future, LCPs will have an increased number of wirelessly interconnected components that are capable of not only pacing and CRT, but also leadless defibrillation therapy~\cite{tjong2017permanent}. They will consist of multiple leadless pacers, heart rate sensors, pressure monitors and can also be combined with other novel devices and therapies. Currently, efforts are being made to combine LCPs and subcutaneous Implantable Cardioverter-Defibrillators (s-ICDs) into a single interconnected ecosystem~\cite{tjong2016combined}. 

Figure~\ref{fig:future_scenario2} shows the basic topology of this second use case. It consists of multiple leadless pacers in the right atrium and both ventricles, a pulmonary artery pressure monitor and an interconnected s-ICD. Given their capabilities to record the heart rate and deliver treatment in the form of electrical impulses, leadless pacers can be considered as hybrid devices, while the blood pressure monitor is a pure sensor. The s-ICD's device generator is used as the central coordinator for the entire LCP-network as it offers the greatest resource capacities. It is not only positioned subcutaneously but also extrathoracicly, making it relatively easy to access for reprogramming and replacing in case of device failure. Since the generator is the network's coordinator and all other nodes directly communicate with it, the network is arranged in a star-topology. For safety and resource purposes it is not assumed that the CCU will directly handle beyond-MBAN communications. Therefore, there is a possibility for the CCU to connect to a wearable or ambient personal device (e.g., mobile phone, smart watch, bedside reader etc.), which connects the network to the Internet to transmit recorded medical data to a clinical server for further processing. The network's topology would in this case remain unchanged, as the personal device is merely a relay node that connects the MBAN to the outside world. If the personal device is dysfunctional or not within the communication range, recorded data is stored in memory until a secure connection to the personal device is established once again. The nodes and the coordinator communicate with each other using intra-body communication. It is assumed that the implanted nodes have enough computational capabilities to encrypt this communication~\cite{siddiqi2019imd,siddiqi2021adding}.

The specifications for this use case are as follows:

\begin{itemize}
    \item \textbf{U2.1:} Consider the moderate resource character of implantable nodes (E2, M2, C2)
    \item \textbf{U2.2:} Support the network's star topology (T1)
    \item \textbf{U2.3:} Guarantee that keys stored on the personal device are only accessible by authorized entities (S5, S11)
    \item \textbf{U2.4:} Dynamically associate/disassociate the personal device with the CCU, as the personal device will not always be in reach (S9)
    \item \textbf{U2.5:} Make sure that only authorized personal devices establish a connection to CCU while ensuring accountability (S5, S6)
    \item \textbf{U2.6:} Encrypt communication between implanted nodes, the CCU and the personal device (S1, S2)
    \item \textbf{U2.7:} Ensure mutual authentication between implanted nodes and the CCU, and between the CCU and the personal device (S4)
    \item \textbf{U2.8:} Support high availability and robustness of the system given the high criticality of its function (S3, S8, S10)
    \item \textbf{U2.9:} Ensure data frames are protected with non-repeating sequence numbers to mitigate the risk of eavesdropping and replay attacks (S1, S2, S7)
\end{itemize}

\subsubsection{Use case 3: Artificial Pancreas}
\label{subsec:scenario_3_artificial_pancreas}

Type 1 diabetes is one of the most common chronic diseases to date. If blood glucose levels remain unmanaged a variety of micro- and macro-vascular diseases, like cardio-vascular or renal failure, limb amputations, vision loss or nerve damage can occur~\cite{bommer2018global}. Despite intensive research, only reactive interventions are available and feasible to date. The most common method is to measure blood sugar by first pricking the finger to retrieve a blood drop. If the blood-sugar level is higher than normal, insulin is manually administered by an insulin pen or a pump. There are two types of issues with this traditional method: (a) Pricking the finger is rather uncomfortable and only gives information about the blood sugar levels at one specific point in time and (b) manually administering insulin is error-prone, ineffective and inconvenient. However, two recent technological advancements in this field have enabled this process to be fully automated. \textit{Continuous glucose monitors} (CGM) can be used to continuously measure real-time values of blood glucose levels. Usually, a CGM consists of a sensor that is placed just beneath the skin to measure blood glucose levels, and a wireless rechargeable transmitter that is fastened on top to send collected data to the outside world. There are already several commercial products, such as the Medtronic Guardian Connect or the Dexcom G6 CGM system, which are both placed on the abdomen. The second advancement is a \textit{closed-loop insulin pump}, which is able to autonomously administer insulin without the need of any user intervention. An extension of that is a system that can administer both insulin and glucagon to also raise blood glucose levels if needed. Such systems are called \textit{bihormonal insulin pumps} or sometimes also \textit{artificial pancreas} since they essentially mimic the pancreas' function. While there is a lot of active research around these types of insulin pumps, no commercial systems exist to date.

\begin{figure}[!t]
\centering
  \includegraphics[scale=.68]{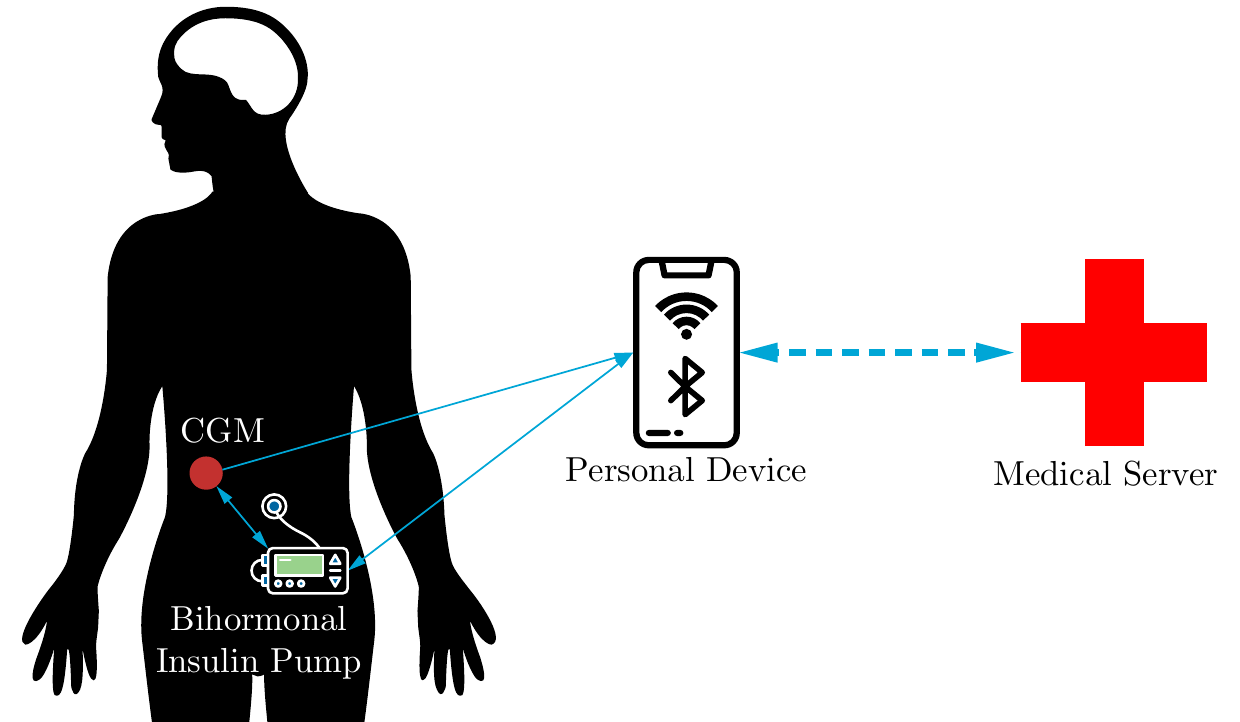}
  \caption{Overview of use case 3: A CGM system recording real-time blood glucose levels that transmits data to a closed-loop bihormonal insulin pump and a personal device for relaying data to a medical server}
  \label{fig:scenario_3_artificial_pancreas}
\end{figure}

In this third use case, the functionality of a CGM and that of an artificial pancreas is combined in order to create an ecosystem to continuously monitor and regulate blood glucose levels (see Figure~\ref{fig:scenario_3_artificial_pancreas}).
The CGM system records real-time blood-sugar levels and transmits the data to a closed-loop bihormonal insulin pump, as well as a personal device (e.g., smartphone or smart watch). Besides the obvious functionality of the pump, it also provides extended memory capacities to store collected medical data if the personal device is not in reach. If the personal device connects at a later point in time the pump transmits the historical and current dosage data, and other stored medical data received from the sensor. It also acts as the network's central coordinator, handling security processes, medium access and power management. The personal device processes the received medical data and displays it, creating the opportunity for the patient to better understand how the body reacts to meals and physical exercise, and to evolve healthy habits. The personal device also acts as a relay node to transmit data to a medical server for further processing and deeper analysis. The network is arranged in a peer-to-peer fashion as the individual nodes need to be connected to each other to exchange data.

The use-case specifications are as follows:

\begin{itemize}
    \item \textbf{U3.1:} Consider the high resource character of implantable nodes (E3, M3, C3)
    \item \textbf{U3.2:} Support the MBAN's peer-to-peer topology (T3)
    \item \textbf{U3.3:} Ensure encryption of all the peer-to-peer connections (S1, S2)
    \item \textbf{U3.4:} Guarantee that only an authenticated and authorized personal device connects to the network (S4, S5, S6)
    \item \textbf{U3.5:} Protect the pump and the CGM from DoS attacks given the importance of their function (S3, S8, S10)
    \item \textbf{U3.6:} Ensure that the data transmitted by the CGM is up-to-date and not tampered with (S2, S7)
    \item \textbf{U3.7:} In case of a battery change the links need to dynamically associate/disassociate with each other (S9)
    \item \textbf{U3.8:} Make sure that keys are safely stored on each node and properly managed by the CCU (S11)
\end{itemize}

\begin{table*}[!t]
    \caption{Summary of the key-features of the three MBAN use cases}
    \centering
    \begin{tabular}{c}
        \includegraphics[trim={3cm 0cm 3cm 0cm},clip,scale=0.6]{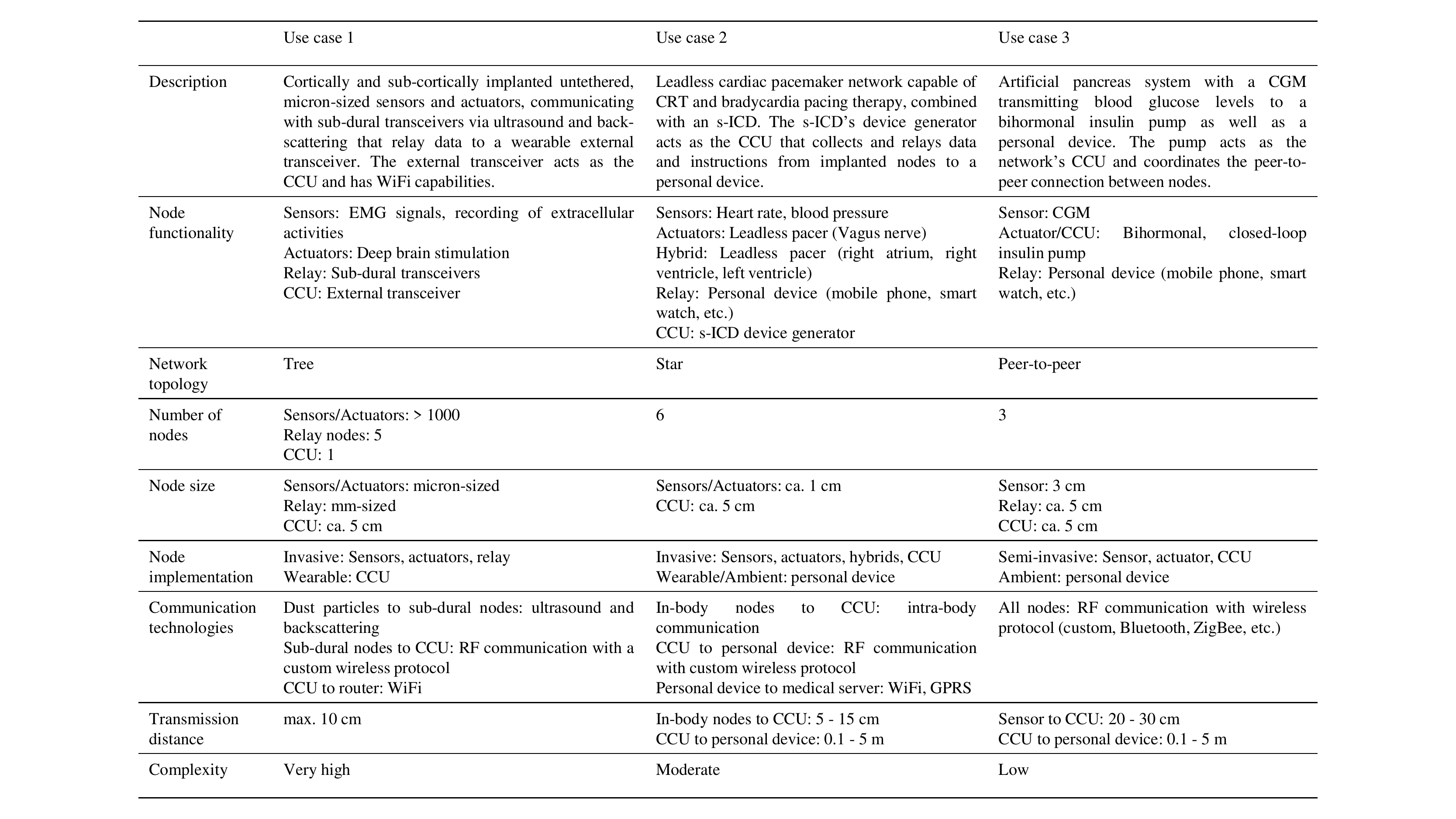}\\
    \end{tabular}
    \label{tab:futuristic_scenarios}
\end{table*}

\begin{table*}[!t]
    \caption{Summary of security(-related) attributes of the three MBAN use cases. A complete coverage of all the defined attributes Sx, Px and Dx is clearly seen.}
    \centering
    \begin{tabular}{c}
        \includegraphics[trim={0cm 3.2cm 0cm 3.2cm},clip,scale=0.51]{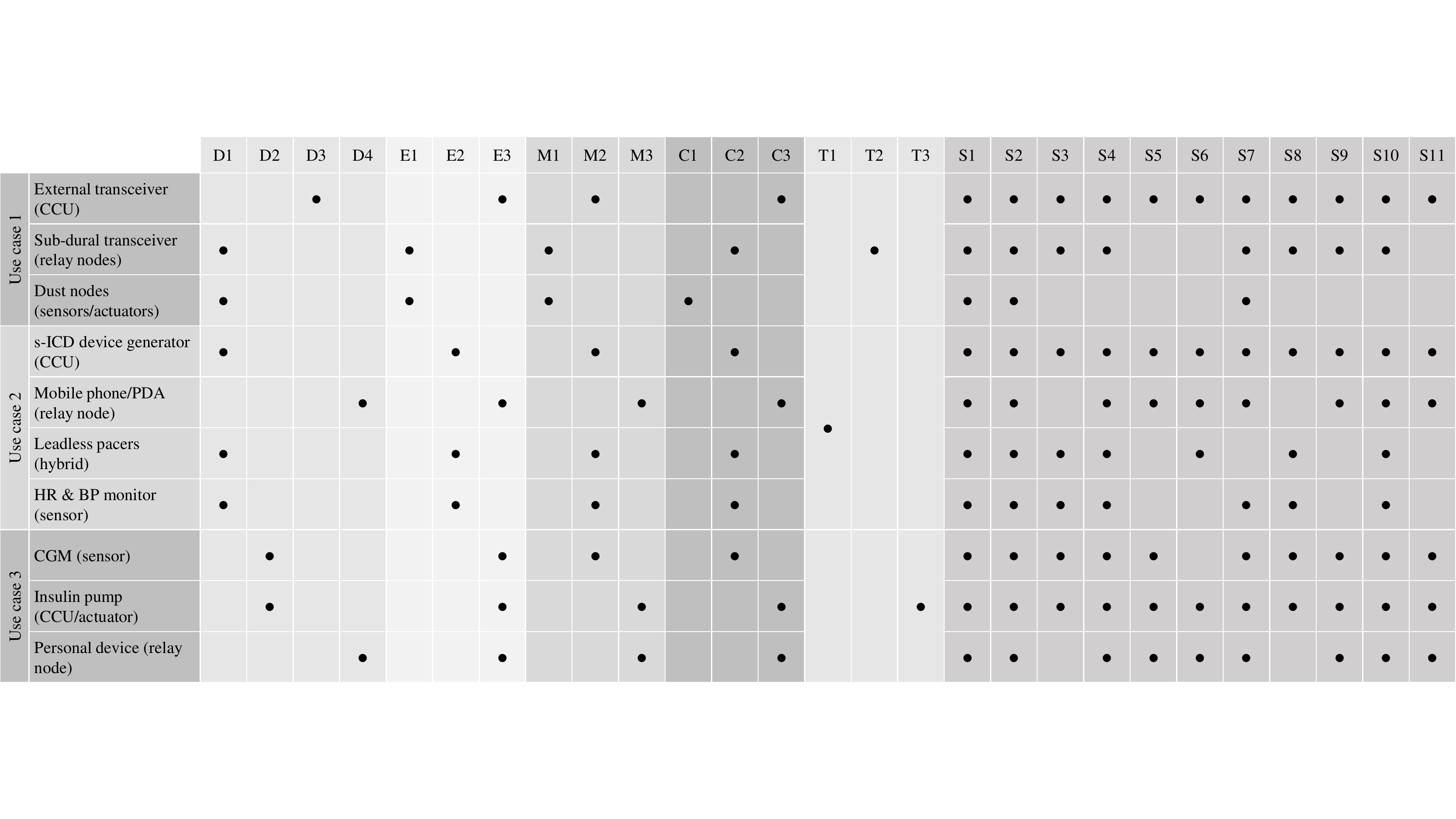}\\
    \end{tabular}
    \label{tab:master_matrix}
\end{table*}

Table~\ref{tab:futuristic_scenarios} summarizes the key features of the use cases introduced in this section. These use cases represent promising future applications that offer a variety of exciting functionalities and treatments. They collectively cover a wide range of possible challenges that the IEEE 802.15.6 standard ought to be able to address.

In Table~\ref{tab:master_matrix}, we connect the range of security(-related)attributes and the specific components (i.e., nodes) of each use case.
This is followed by connecting the use-case specifications to the previous attributes (Sx and Px) by using a similar matrix shown in Table~\ref{tab:application_specific_matrix}.
By inspecting these tables, we can confirm the suitability of our selection: when combined, the three MBAN use cases cover all the different attributes (Sx, Px and Dx) defined in our assessment methodology, meaning that no security aspect of a future MBAN design will remain unaddressed by our assessment methodology. Though our methodology takes a heuristic, bottom-up approach to assessing the standard, it is considered the most practical approach to a vaguely defined and -- specifications-wise -- open-ended field such as MBANs are at the moment. The power of this approach is that it can yield recommendations for the standard immediately, it maintains a low handling complexity and it allows for any number of extensions -- be it in adding more use cases or more security(-related) attributes in the future -- without breaking.

\subsection{Putting the IEEE standard to the test}
\label{sec:putting_the_IEEE_802.15.6_standard_to_the_test}

The assessment will follow a clear and structured procedure, where the use-case specifications are reflected against the standard one by one. Eventual discrepancies will be documented and inserted in a color-coded matrix (Table~\ref{tab:application_specific_matrix}).

\subsubsection{Use case 1: Neural Dust}

\quad \textbf{U1.1:} The IEEE 802.15.6 standard claims to be designed for low-power devices, supporting data rates up to 10 Mbps, while keeping the specific absorption rate (SAR) to a minimum. At the MAC level, the standard provides recommendations for power management of nodes, in which they have the ability to enter hibernation or sleep mode for energy-saving purposes. Although sub-dural transceivers and implanted dust nodes are of passive nature, this is still relevant for the external transceiver as it is battery-powered and the main source of energy for the entire intra-MBAN network. This means that if the external transceiver is not powering the sub-dural transceivers via RF power transfer, the underlying nodes are always in a hibernation/sleep state. At the PHY layer, the standard offers the possibility to employ UWB communication, which is a proven technology for ultra-low power devices. Specifically, UWB-RFID is a promising technology for the communication between the external and sub-dural transceivers. However, it is still not clear if the mm-sized sub-dural transceivers and the even smaller dust nodes have enough resources to handle the standard's protocols.

\textbf{U1.2:} The standard dictates the network topology by limiting the number of hubs in an MBAN to a single one. The topology discussed in the standard is a star with the possibility to employ a two-hop star extension, which is needed for this use case.

\textbf{U1.3:} The maximum number of nodes supported by the standard is specified in the parameter \textit{mMaxBANSize}, which is equal to 64. Considering that a neural-dust application must support thousands of individual dust nodes, this number is not sufficient. The standard does not specify where this limitation comes from. It can only be assumed that computational and memory resources of the CCU might be seen as limited, i.e., they cannot support a higher number of nodes.

\textbf{U1.4:} To see if the standard fulfills this specification, certain assumptions about the computational capabilities of the mm-sized sub-dural transceivers have to be made. Although the transceiver's form factor is in a range where Moore's law limits their computational capabilities, it is assumed that in contrast to the micron-sized dust nodes they still have sufficient processing power to handle the cryptographic algorithms suggested by the standard. Hence, this specification can be considered as fulfilled.

\textbf{U1.5:} The standard mentions mutual authentication for two of the five association protocols, i.e., protocols I (pre-shared MK) and IV (display authenticated). Protocol IV, however, is not valid in this specific use case as the CCU does not have a display for authenticating the 5-digit number. Protocol I ensures mutual authentication by using the pre-shared, readily activated MK, while simultaneously initiating the PTK creation procedure. The exact workings of the mutual authentication procedure are not mentioned in the standard. For the remaining three protocols, the standard does not specifically mention mutual authentication, which is why this design specification is only partly fulfilled.

\textbf{U1.6:} The standard uses a variety of keys, e.g., MK, PTK, GTK etc. to handle association and authentication. Thereby, secure key management is crucial to mitigate the risk of the keys being compromised. There are detailed protocols to generate, distribute, refresh and revoke keys employed by the standard.

\textbf{U1.7:} In case of a node failure, a protocol must be in place to communicate the occurrence of a failure to the hub and subsequently to the medical server, where actions can be decided. Failure of individual dust nodes might be tolerable to a certain degree without having to initiate intervention procedures. Thereby, the standard must account for changes in the network topology and computational overhead of the network's nodes. The standard does not specify what happens if a node disconnects in case of a failure. In fact, node failure is not mentioned at all. Additionally, the standard does not include measures to improve dependability in terms of security.

\textbf{U1.8:} To ensure message freshness and protect against replay attacks, the standard implements \textit{low-} and \textit{high-order security sequence numbers}. If a frame is secured with the same PTK or GTK, the value of the low-order number increments by one. This is also true for re-transmission of previous frames. If a node or hub receives a frame that causes the high-order sequence number to wrap around zero, it will be discarded. The same will happen if the value of the low-order sequence number of the current frame is not higher than that of the previous frame.

\textbf{U1.9:} The security paradigm mentioned in the standard offers the possibility to authenticate control type frames during security-frame exchanges between external and sub-dural transceivers. As mentioned in Section~\ref{sec:IEEE_802.15.6_security}, the SSS can be used to specify if either security level 1 or 2 shall be employed. The main tool of authentication is the Cipher-based Message Authentication Code (CMAC), as specified in the NIST Special Publication 800-38B, which is used to compute Key Message Authentication Codes (KMAC) and the MK. CMACs are considered to be energy efficient and memory saving if the authenticated messages are up to two blocks long~\cite{simplicio2013survey}. However, the standard fails to introduce a mechanism (e.g., access-control lists etc.) for addressing authorization. Hence, this specification is not fulfilled.

\subsubsection{Use case 2: Leadless Cardiac Pacemaker}

\quad \textbf{U2.1:} For this use case, the low-power measures and protocols offered by the standard are considerate of the application-specific restrictions. The individual implanted nodes have sufficient capabilities to handle computations (such as encryption). Additionally, the power-management options provided by the standard (i.e., nodes being able to hibernate/sleep) aid in saving energy.

\textbf{U2.2:} As already mentioned in the previous use case, the standard supports networks with a star topology, meaning this specification is satisfied.

\textbf{U2.3:} As already established in the previous use case, the standard does not offer any recommendations on how to securely store generated keys. In this use case, the CCU is implanted beneath the patient's skin, enabling the device a certain degree of inherent security. However, this is not the case for keys stored on the personal device. Here, very strict security controls have to be in place to ensure that keys are not stolen.

\textbf{U2.4:} The association and disassociation frames specified in the standard are employed to regulate connections between a hub and accompanying nodes. Based on the \textit{connection-status} field, connection requests are categorized for deciding whether a connection can be established or terminated. While this is sufficient in most situations, the standard does not treat cases where nodes unexpectedly or abruptly terminate an already established connection, e.g., through failure or an empty battery, thus, not having an opportunity to send a disconnection frame. It does, however, support dynamic association by sending connection request frames if the personal device is in reach.

\textbf{U2.5:} The standard has protocols in place to ensure that a hub-node pair follows a specific security hierarchy when initially establishing a mutual connection. However, if a node follows the specified processes to generate an SSS, a valid public key and other security frames, there is nothing stopping a malicious actor to connect this node to the network, as no access-control list or other mechanism to track legitimate nodes is discussed by the standard. Hence, this specification is not satisfied.

\textbf{U2.6:} The resource-constrained nature of implantable nodes allows sufficient processing power to handle the cryptographic algorithms suggested by the standard.

\textbf{U2.7:} As already discussed when assessing the first use case, the standard only mentions mutual authentication specifically for the association protocols I (pre-shared MK) and IV (display authenticated). Once a PTK is established between a hub-node pair, the origin (i.e., the sender's address) is corroborated using the CCM mode of AES whenever a message is transmitted. At the control-frame level, frames are also authenticated using AES in CCM mode as well as Message Integrity Codes (MICs).

\textbf{U2.8:} Although the standard provides measures to increase robustness at the bit-level (e.g., bit interleaving prior to modulation) and at the PHY and frequency band level (e.g., UWB, FM-UWB), it does not discuss robustness against security threats. There are no controls in place to detect or actively prevent intrusion in the network, or protect against DoS attacks, which can have a detrimental effect on the system's availability. Furthermore, the standard does not discuss the threat landscape or possible attack scenarios in order to improve availability and robustness, which is why this specification is considered as not satisfied.

\textbf{U2.9:} As already mentioned in the previous use case, the security sequence number used in the standard aids in protecting messages from replay attacks and ensures data freshness. Hence, this specification can be considered as satisfied.

\subsubsection{Use case 3: Artificial Pancreas}

\quad \textbf{U3.1:} Given the high resource capabilities of the devices used in use case 3, the protocols of the standard can be easily computed and handled in terms of resource allocation. Hence, this specification is fulfilled by the standard.

\textbf{U3.2:} The standard only mentions the star topology for medium access. Additionally, a two-hop star extension can be employed. The peer-to-peer topology of use case 3's network is not discussed, which is why this specification is not fulfilled.

\textbf{U3.3:} The high resource availability of this use case's MBAN nodes allows sufficient processing power to handle the cryptographic algorithms suggested by the standard.

\textbf{U3.4:} Following the reasoning of U1.9, this specification is not fulfilled.

\textbf{U3.5:} The main focus of the standard in terms of robustness lies in reducing transmission errors and interference. There is no mention of DoS protection or any other controls to harden the system against security attacks, hence, leaving it to the designer to implement dependability measures.

\textbf{U3.6:} Following the reasoning of U2.9, this specification can be considered as satisfied.

\textbf{U3.7:} Following the reasoning of U2.4, this specification can be considered as partially satisfied.

\textbf{U3.8:} The standard successfully implements the following important aspects of key management: generation, refreshing, agreement, distribution and revocation. However, as discussed in Section~\ref{subsec:security_vulnerabilities_and_weaknesses}, certain vulnerabilities in the standard's association protocols have been discovered, which pose a major issue for secure key management. Furthermore, the standard does not mention how keys should be secured when at rest, opening possibilities for attacks on the physical nodes.

\begin{table*}[!t]
    \caption{Correlation matrix of the various use-case specifications (Ux.x) with the security and physical attributes (Sx and Px). Moreover, the matrix identifies the areas where the standard does not fulfill Ux.x and, hence, Sx and Px. (Red: The standard does not satisfy this specification; Yellow: The standard partly satisfies this specification; Green: The standard satisfies this specification).}
    \centering
    \begin{tabular}{c}
        \includegraphics[trim={3.9cm 0.4cm 0.0cm 0.8cm},clip,scale=0.67]{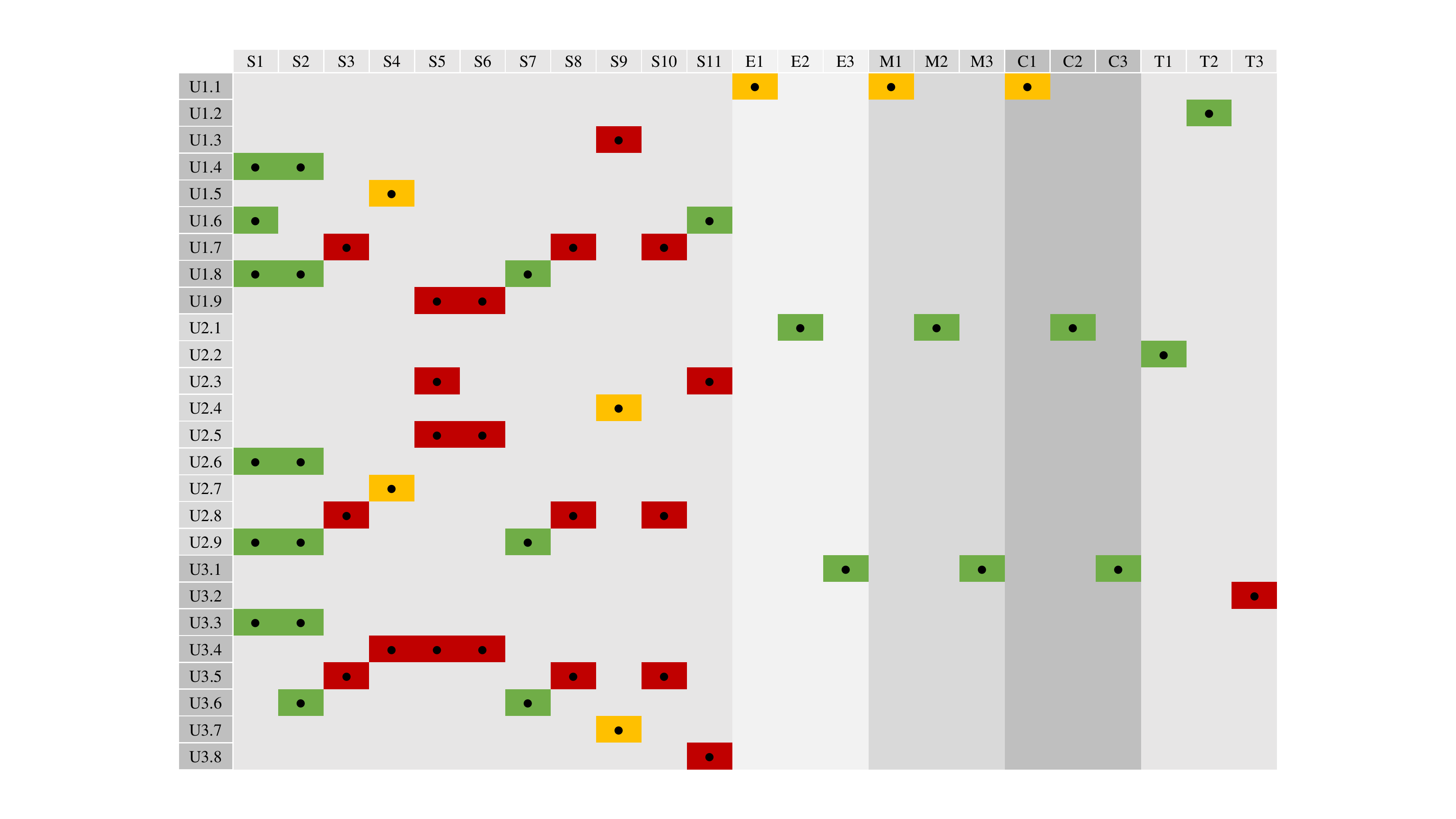}\\
    \end{tabular}
    \label{tab:application_specific_matrix}
\end{table*}

Table~\ref{tab:application_specific_matrix} summarizes the standard's fulfillment of the specifications of the MBAN use cases. The table is color-coded to indicate to what degree the standard is fulfilling each Sx and Px.
We can see several shortcomings that have been identified by following the structured analysis procedure. In the next step, the findings and the related recommendations will be summarised.

\subsection{Recommendations for improving IEEE 802.15.6}

As already discussed in Section~\ref{sec:introducing_the_assessment_procedure}, the recommendations of our analysis will be clustered into three main categories, namely physical, organizational and security recommendations. Table~\ref{tab:findings_connection} gives a general overview of how the recommendations connect to the use-case specifications.
Even though the standard focuses on PHY and MAC layers, we have still included a few recommendations that are not limited to just these layers. This is inline with the holistic approach of Section~\ref{chap:mban_security_requirements_and_threat_landscape}.

\begin{table*}[!t]
    \caption{Connection between the recommendations for improving the standard in terms of security and the defined use-case specifications (Ux.x).
    \ifnum\hltext>0
    \hl{The recommendations that do not pertain to PHY and MAC layers are colored in blue.}
    \else
    The recommendations that do not solely pertain to PHY and MAC layers are colored in blue.
    \fi
    }
    \centering
    \begin{tabular}{c}
        \includegraphics[trim={2.9cm 0cm 0cm 0cm},clip,scale=0.62]{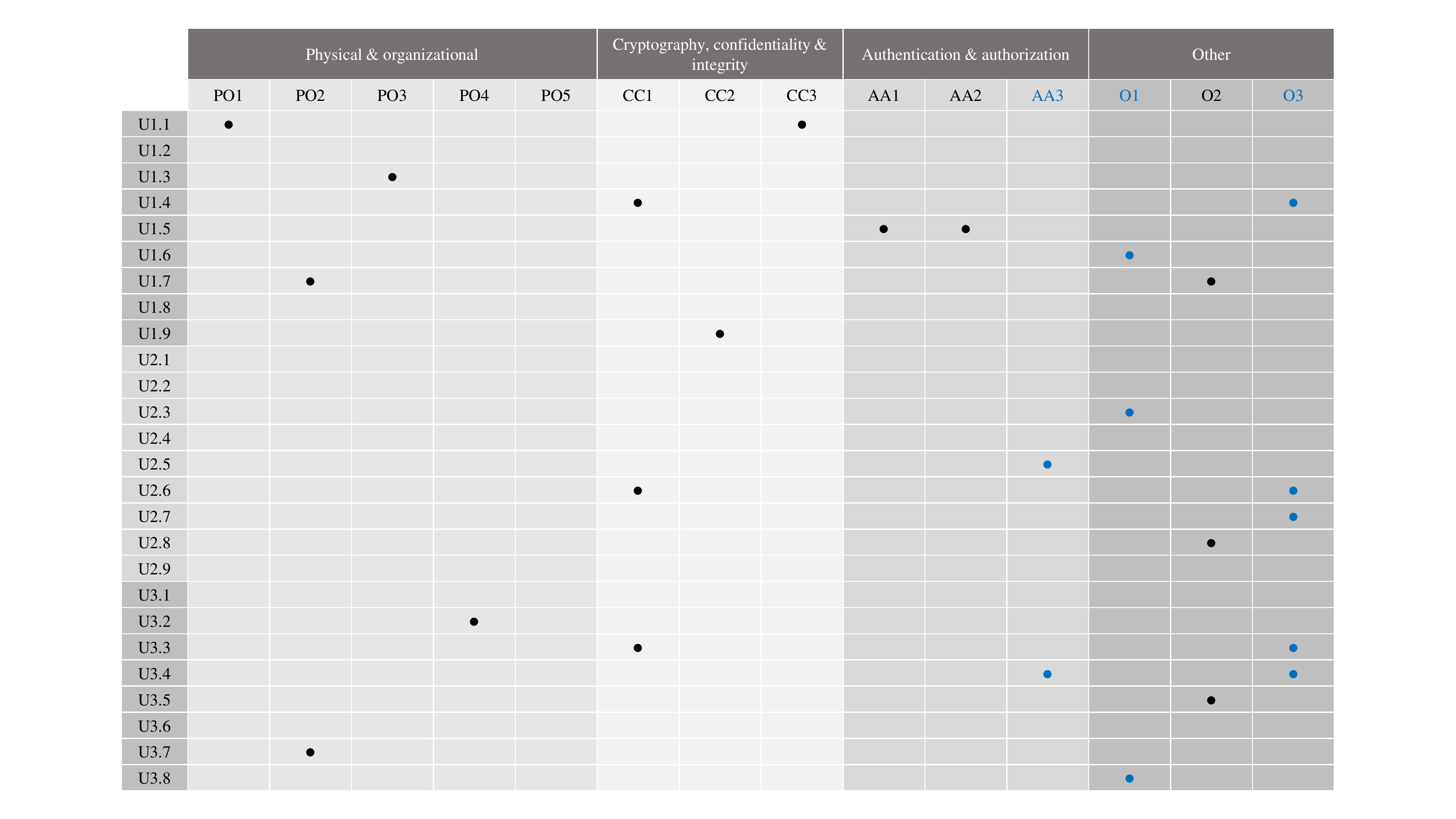}\\
    \end{tabular}
    \label{tab:findings_connection}
\end{table*}

\subsubsection{Physical and organizational recommendations}

\quad \textbf{PO1:} 
Although the standard is designed for ultra-low power devices, it does not consider passive devices that have no computational capabilities at all.
Therefore, a specific section on how to include passive devices and handle their specific requirements has to be added in the following chapters
\ifnum\hltext>0
\hl{of~\mbox{\cite{ieee_standard}}}:
\else
of~\cite{ieee_standard}:
\fi
    \begin{itemize}
        \item \textit{4. General framework elements:} This chapter explains the framework elements within the defined scope, which does not include passive devices.
        \item \textit{10. Human body communications PHY specification:} Since most of implantable passive devices rely on human body communication, a section needs to be added explaining in detail the communication of passive devices at the PHY level.
    \end{itemize}

\textbf{PO2:}
While there is a whole list of possible reasons why an incoming connection request gets rejected, there is no discussion on what happens if a node abruptly disconnects due to device failure or DoS.

Section \textit{6.4.3 Node disconnection} specifies that if a node receives a Connection Assignment frame indicating that a connection request was rejected for some reason, the node sends a Disconnection frame. The possible reasons for a rejected connection request are listed in \textit{Table 12 - Connection status field encoding}. An entry needs to be added to this table stating that a connection request was rejected because the device is not reachable. Whether this is due to device failure or a DoS attack does not matter in this context. In addition to that, an explanation on what happens if the node can no longer communicate with the previously connected device needs to be added within Section \textit{6.4.3}.

\textbf{PO3:}
The standard only supports a maximal number of nodes of up to 64 (\textit{mMaxBANSize}). 
Furthermore, it is not mentioned where this limit comes from.

An explanation of this limitation needs to be added in Chapter \textit{4.2 Network topology} or as an addition to \textit{Table 24 - MAC sublayer parameters}. More importantly, this limitation is not sufficient for future applications (e.g., Neural Dust).

\textbf{PO4:}
The only network topology introduced in the standard is a star with the possibility of a two-hop star extension. Peer-to-peer topologies and ad-hoc connections are not discussed.

As a result, the standard needs to include the peer-to-peer topology in addition to the star network introduced in Chapter \textit{4.2 Network topology}. Additionally, the standard needs to either extended the existing protocols and procedures to support multiple topologies, or new protocols need to be introduced.

\textbf{PO5:}
As per the standard, the BAN must have one and only one hub. Networks that require more than one hub (for instance, as a passive back-up) are therefore not covered by the standard. Moreover, the standard needs to consider the scenario when a new hub gets added to the network or the current hub is replaced. Therefore, the restriction in the first paragraph of Chapter \textit{4.2 Network topology} needs to be lifted and the introduced topology needs to be extended.

\subsubsection{Cryptography, confidentiality and integrity recommendations}

\quad \textbf{CC1:}
The AES and Camellia ciphers used for encryption are only issued with a key size of 128 bits. However, there are already more secure, bigger key-sizes available, which should be employed instead whenever possible.

In \textit{Table 7 - Cipher Function field encoding} of Chapter \textit{4.3.2.3.4 Cipher function}, additional entries for the 256-bit versions of both forward cipher functions should be added. Furthermore, Chapter \textit{7. Security services} needs to mention the possibility of using larger key sizes.

\textbf{CC2:}
The standard uses CMACs according to the NIST special publication 800-38B, which are energy-efficient and memory-saving for messages up to 2 blocks long~\cite{simplicio2013survey}. However, alternatives for longer messages are not discussed.

In chapters \textit{7.1 Security association and disassociation} and \textit{7.2 PTK creation and GTK distribution} the definitions of the CMAC used to compute KMACs needs to be extended to include more energy efficient CMAC alternatives for messages that are longer than 2 blocks.

\textbf{CC3:}
For applications employing devices that do not have sufficient processing power to compute complex security algorithms, the standard does not provide any guidelines.

In Chapter \textit{7. Security services}, a paragraph needs to be added stating that for devices that cannot support conventional cryptography, the developer needs to ensure the inherent security of the link (security by design). For example, in Neural Dust, a MHz-range ultrasound channel can be employed to achieve inherently-secure communication.

\subsubsection{Authentication and authorization recommendations}

\quad \textbf{AA1:}
In protocol I, the standard mentions a mutual authentication procedure based on the pre-shared MK. However, it does not specify the inner workings of this procedure.

In Chapter \textit{7.1.1 Master key pre-shared association}, it is mentioned that the hub and the node perform mutual authentication with each other, while simultaneously advancing to the PTK creation procedure. Here, a paragraph explaining the inner workings of the mutual authentication procedure has to be added.

\textbf{AA2:}
For the public-key hidden association protocol (protocol III) the standard fails to specifically mention mutual authentication.

In Chapter \textit{7.1.3 Public key hidden association}, the standard needs to specifically mention that mutual authentication should be guaranteed in case the association procedure is successful.

\textbf{AA3:}
The standard does not provide or recommend an access-control mechanism.
Any node with the correct formal requirements (e.g., frame structure, SSS etc.) can establish a connection with the hub.

In Chapter \textit{7. Security services}, the standard needs to introduce and discuss a means to store and track created and established keys to guarantee that only registered devices can enter the network. The concept of an access control list -- managed by the hub -- can be used to keep track of authorization of individual nodes.

\subsubsection{Other recommendations}

\quad \textbf{O1:}
Although the standard offers various protocols for secure key management, it fails to discuss how security keys should be stored on devices that allow physical access by an attacker.

In Chapter \textit{7. Security services}, a section on best practices and recommendations to securely store security keys when at rest needs to be included. This is especially important for nodes that are physically accessible, e.g., semi-invasive, wearable and ambient nodes.

\textbf{O2:}
The standard does not deploy sufficient measures to protect the network against DoS attacks, nor to improve the overall dependability of the system.
In Chapter \textit{7. Security services}, best practices and recommendations on how to harden a BAN against such attacks need to be introduced in a separate sub-chapter.

\textbf{O3:}
The security of the standard mainly revolves around authentication and encryption. Broader topics, such as the overall threat surface and possible attack scenarios are not part of the discussion. Therefore, the standard needs to include recommendations on how to effectively reduce the threat surface of an MBAN (in Chapter \textit{7. Security services}).

%% file: 08_Conclusion.tex
\section{Conclusions}
\label{chap:conclusion}

This paper focused on improving the security of future MBAN applications by assessing the security posture of the IEEE 802.15.6 standard, which aims to govern such networks. Therefore, a structured methodology to assess this standard was introduced. The main difficulty in such an analysis is guaranteeing an exhaustive study of all potential MBAN configurations, covering all of the necessary dimensions. Since such an approach is impractical and often intractable, our heuristic methodology proposes an assessment strategy whereby a few MBAN use cases are studied as a proxy for the whole field. The challenge lies in selecting highly representative and relevant use cases, while the strength of our approach is that it is an iterative and expandable one; that is, it can accommodate more use cases and more security vulnerabilities. Besides, as device and network security are dependent on node types, the assessment needed to encompass more than just the scope of security itself. Therefore, not only security attributes, but also a number of physical attributes that represent distinct node types were taken into consideration. These attributes were then cast against the design specifications of three realistic and mutually exclusive use cases. The assessment ultimately resulted in various concrete recommendations for improving the IEEE 802.15.6 security, which will be communicated to the 802.15 working group. It should be noted that our assessment procedure is not specific to the particular standard and can be extended to other similar standards and applications. Moreover, many security issues applicable to general MBAN applications, and not limited to only IEEE 802.15.6, were raised in this paper.